\documentclass[preprint,11pt,numbered, sn-mathphys-num]{sn-jnl}
\usepackage[utf8]{inputenc}
\usepackage[sort&compress]{natbib}
\usepackage{fancyhdr}
\usepackage{amssymb,amstext,amsmath, amsthm}
\usepackage{pifont}
\usepackage{booktabs}
\usepackage{bm}
\usepackage{graphicx}
\usepackage{notoccite}
\usepackage{color}
\usepackage[labelformat=simple]{subcaption}
\usepackage{csquotes}
\usepackage{longtable}
\usepackage{array}
\newcolumntype{L}[1]{>{\raggedright\let\newline\\\arraybackslash\hspace{0pt}}p{#1}}
\usepackage{svg}
\usepackage[ruled,longend,linesnumbered]{algorithm2e}
\usepackage{hhline}
\usepackage{multirow}
\newcommand{\boldface}[1]{\boldsymbol{#1}}  
\newcommand{\bfa}{\boldface{a}}

\newcommand{\bfe}{\boldface{e}}

\newcommand{\bfn}{\boldface{n}}

\newcommand{\bfs}{\boldface{s}}

\newcommand{\bfu}{\boldface{u}}

\newcommand{\bfF}{\boldface{F}}

\newcommand{\bfX}{\boldface{X}}
\newcommand{\bfY}{\boldface{Y}}

\newcommand{\bflambda}{\boldsymbol{\lambda}}

\newcommand{\bfvarphi}{\boldsymbol{\varphi}}

\newcommand{\calA}{\mathcal{A}}
\newcommand{\calB}{\mathcal{B}}

\newcommand{\calE}{\mathcal{E}}

\newcommand{\calI}{\mathcal{I}}

\newcommand{\calL}{\mathcal{L}}

\newcommand{\calN}{\mathcal{N}}

\newcommand{\calS}{\mathcal{S}}

\newcommand{\half}{\frac{1}{2}}

\newcommand{\Rset}{\mathbb{R}}

\newlength{\boxwidth}
\def\btheorem{\begin{theorem}}
\def\etheorem{\end{theorem}}
\def\blemma{\begin{lemma}}
\def\elemma{\end{lemma}}
\def\bproposition{\begin{proposition}}
\def\eproposition{\end{proposition}}
\def\bcorollary{\begin{corollary}}
\def\ecorollary{\end{corollary}}
\def\bdefinition{\begin{definition}}
\def\edefinition{\end{definition}}
\def\bexample{\begin{example}}
\def\eexample{\end{example}}
\def\bremark{\begin{remark}}
\def\eremark{\end{remark}}

\DeclareMathOperator{\argmin}{{arg\,min}}

\newcommand{\be}{\begin{equation*}}
\newcommand{\ee}{\end{equation*}}

\newcommand{\beq}{\begin{eqnarray*}}
\newcommand{\eeq}{\end{eqnarray*}}
\newcommand{\bem}{\begin{multline}}
\newcommand{\eem}{\end{multline}}
\newcommand{\ba}{\begin{align*}}
\newcommand{\ea}{\end{align*}}

\renewcommand{\figurename}{Figure}
\renewcommand{\tablename}{Table}

\def\CC{{C\nolinebreak[4]\hspace{-.05em}\raisebox{.45ex}{\tiny\bf ++}}}
\begin{document}

\title[Article Title]{A mixed-order quasicontinuum approach for beam-based architected materials with application to fracture}

\author[1]{\fnm{Kevin} \sur{Kraschewski}}\email{kkraschewski@ethz.ch}

\author[2]{\fnm{Gregory P.} \sur{Phlipot}}\email{gphlipot@alumni.caltech.edu}
\author*[1]{\fnm{Dennis M.} \sur{Kochmann}}\email{dmk@ethz.ch}

\affil*[1]{\orgdiv{Mechanics \& Materials Lab, Department of Mechanical and Process Engineering}, \orgname{ETH Z\"urich}, \orgaddress{\street{Leonhardstr.~21}, \city{Zurich}, \postcode{8092}, \country{Switzerland}}}
\affil[2]{\orgdiv{Divison of Engineering and Applied Science}, \orgname{California Institute of Technology}, \orgaddress{\street{1200 E. California Blvd.}, \city{MC 155-44 Pasadena}, \postcode{91125-2100}, \state{CA}, \country{USA}}}

\abstract{
Predicting the mechanics of large structural networks, such as beam-based architected materials, requires a multiscale computational strategy that preserves information about the discrete structure while being applicable to large assemblies of struts. Especially the fracture properties of such beam lattices necessitate a two-scale modeling strategy, since the fracture toughness depends on discrete beam failure events, while the application of remote loads requires large simulation domains. As classical homogenization techniques fail in the absence of a separation of scales at the crack tip, we present a concurrent multiscale technique: a fully-nonlocal quasicontinuum (QC) multi-lattice formulation for beam networks, based on a conforming mesh. Like the original atomistic QC formulation, we maintain discrete resolution where needed (such as around a crack tip) while efficiently coarse-graining in the remaining simulation domain. A key challenge is a suitable model in the coarse-grained domain, where classical QC uses affine interpolations. This formulation fails in bending-dominated lattices, as it overconstrains the lattice by preventing bending without stretching of beams. Therefore, we here present a beam QC formulation based on mixed-order interpolation in the coarse-grained region -- combining the efficiency of linear interpolation where possible with the accuracy advantages of quadratic interpolation where needed. This results in a powerful computational framework, which, as we demonstrate through our validation and benchmark examples, overcomes the deficiencies of previous QC formulations and enables, e.g., the prediction of the fracture toughness and the diverse nature of stress distributions of stretching- and bending-dominated beam lattices in two and three dimensions.}

\keywords{Multiscale Modeling; \sep Quasicontinuum; \sep Fracture; \sep Finite Element Method; \sep Structure}

\maketitle
\section{Introduction}
\label{sec:sample1}
Predicting the failure mechanisms of low-density cellular solids, from random fiber networks to periodic architected materials (or metamaterials), has been a challenge for computational mechanics. Especially for periodic architected materials, whose stiffness \citep{meza_reexamining_2017,zheng_ultralight_2014}, wave dispersion \citep{bilal_architected_2018,zelhofer_acoustic_2017}, viscoelastic \citep{glaesener_continuum_2020} and plastic \citep{valdevit_compressive_2013, bauer_additive_2019} deformation behavior, among others, have been studied and simulated extensively \citep{kochmann_multiscale_2019}, the fracture toughness has remained a subject of study and a source of surprise---despite more than a century of fracture mechanics \citep{anderson_fracture_2017}. A key difference between, e.g., beam-based architected materials and classical homogeneous solids is the discrete nature of failure in the former. Even for brittle base materials (which are linear elastic up to failure), this gives rise to significant differences in the fracture behavior compared to continuous solids. While the latter are described well by linear elastic fracture mechanics \citep{griffith_vi_1921,irwin_analysis_1957}, the discrete network fails by breaking individual beams \citep{huang_fracture_1991,gibson_cellular_1999}. As a consequence, instead of showing the classical stress and strain singularities at the crack tip, the load is distributed onto the beams close to the crack tip \citep{schmidt_ductile_2001,fleck_micro-architectured_2010}, so that fracture toughness is defined not through the scaling of the stresses at the crack tip but via the simple question: when does the first beam fail \citep{romijn_fracture_2007}.

Although the fracture toughness of linear elastic beam lattices has been studied for more than two decades, including the works by \citet{chen_fracture_1998, schmidt_ductile_2001, fleck_damage_2007, lipperman_fracture_2007, tankasala_2013_2015,omasta_fracture_2017}, it was shown only recently \citep{shaikeea_toughness_2022} that the common definition of the fracture toughness $K_{IC}$ may be insufficient to describe the fracture of three-dimensional (3D) beam-based architected materials. In their investigation of an octet-lattice consisting of about 10~million unit cells (UCs) with an elastic-brittle constituent material, it was found by \citet{shaikeea_toughness_2022} that the $T$-stresses (i.e., non-singular stresses acting parallel to the crack-edge) can have a significant impact in 3D and that the size of the lattice (in-plane and out-of-plane with respect to the mode-I scenario) affects the fracture behavior. Moreover, their studies, which used fully-resolved finite element (FE) simulations, highlighted the need for advanced computational techniques to efficiently simulate large assemblies of UCs, so that computational power ceases to be a bottleneck for understanding the effective fracture properties of architected materials. Even in two dimensions (2D), the fracture toughness is a complex property of architected materials, which---even for elastically isotropic lattices---may depend strongly on the orientation of the crack (relative to the structural architecture) \citep{lipperman_fracture_2007}. Fracture toughness is, of course, only the tip of the iceberg---considering, e.g., the fatigue of hip stem replacements made of architected materials \citep{kolken_mechanisms_2022}, which sustain around 2~million load cycles per year. All of these applications have one aspect in common: they require efficient computational tools that provide locally the accuracy of a fully-resolved, discrete calculation.

Simulating the effective, macroscopic response of large beam networks by fully-resolved FE calculations (based on continuum elements \citep{kaur_3d_2017} or more efficient beam representations \citep{shaikeea_toughness_2022} and reduced-order models \citep{rimoli_reduced-order_2018}) is computationally expensive, so that multiscale techniques become the method of choice \citep{kochmann_multiscale_2019}. Classical homogenization methods have successfully predicted the linear and nonlinear elastic \citep{deshpande_effective_2001,glaesener_continuum_2019,glaesener_continuum_2020} and plastic behavior \citep{mohr_mechanism-based_2005}. Yet, they rely on the assumption of a separation of scales. If this assumption brakes down, e.g., in the case of fracture, where the fields of interest near the crack tip vary on the same order as the size of individual beams, such methods fail. This is where discrete-to-continuum coupling techniques become a promising alternative.

One popular discrete-to-continuum technique is the \textit{quasicontinuum} (QC) method \citep{tadmor_quasicontinuum_1996}, which was originally developed for atomistic lattices. It maintains full resolution where needed (e.g., around a crack tip), while efficiently coarse-graining in regions away from the crack. In the coarse-grained region, FE interpolation is used to infer from the vertices of the coarse elements the positions (and momenta) of each and every atom within an element. This strategy was extended to periodic truss lattices \citep{beex_quasicontinuum_2011}---atoms being replaced by truss junctions, and atomic potentials being replaced by beam strain energies. As truss lattices tend to be more complex to homogenize in the continuum regions (owing to complex multi-lattice unit cells), the type of \textit{interpolation} in those regions plays an essential role. The initial model of \citet{beex_quasicontinuum_2011} was restricted to 2D and spring-like struts without rotational degrees of freedom, and an \textit{affine interpolation} in the coarse-grained regions (like in the original atomistic QC work). \citet{beex_quasicontinuum-based_2014} extended the method to Euler-Bernoulli beams, using conforming and non-conforming meshes for the coarse-grained regions as well as \textit{higher-order interpolation}. They concluded that a non-conforming triangulation with cubic interpolation for nodal displacements and quadratic interpolation for the rotational degrees of freedom (DOFs) in the coarse regions as well as a linear interpolation of both types of DOFs in the fully-resolved region showed best results. \citet{rokos_adaptive_2017} presented an adaptive refinement strategy to accurately capture crack-propagation in a 2D lattice, only considering translational degrees of freedom (DOFs) and \textit{affine interpolation} for simple nearest-neighbor interactions \citep{beex_higher-order_2015}. Mesh coarsening, which allows to not only adaptively refine ahead of a propagating crack but also to coarsen once the crack has passed, can further reduce computational expenses \citep{rokos_extended_2017}. With the exception of \citet{beex_quasicontinuum-based_2014}, these prior works all focused on \textit{affine interpolation} and (linear or nonlinear) spring models, further on simple Bravais lattices in 2D. 

\citet{phlipot_quasicontinuum_2019} extended the truss QC methodology to multi-lattices, i.e., to more complex periodic networks composed of multiple Bravais lattices shifted relative to each other. This was an important step towards capturing non-affine deformation within coarse-grained unit cells, while retaining affine interpolation of the DOFs. In addition, they presented 3D examples and generalized the beam model to account for large rotations. In a similar approach, \citet{chen_generalized_2020} used multiple beam elements per strut and a plastic hinge model, which mimics varying strut diameters and other imperfections. The more recent works of \citet{chen_refinement_2021} and \citet{chen_adaptive_2022} focused on improving the adaptive refinement algorithm and the incorporation of plasticity models, respectively. Despite the multi-lattice formulation, both \citet{phlipot_quasicontinuum_2019} and \citet{chen_generalized_2020} reported inaccurate predictions for bending-dominated lattices, when applying an affine interpolation scheme in coarse-grained regions. Both theorized that the affine interpolation overconstrains the lattice, which effectively prevents bending of beam members without affinely stretching the unit cell---also referred to as \emph{stretch locking} \citep{phlipot_quasicontinuum_2019}. As a remedy, \citet{chen_refinement_2021} used an improved refinement algorithm, which ensured full resolution in those regions with non-uniform deformation. This, however, significantly reduced the efficiency, since errors below 1\% were reported to require up to $93.78\%$ of all (fully-resolved) DOFs, which implies prohibitive computational expenses.

Here, we introduce a truss QC framework that integrates several previously disjointed techniques into a versatile multiscale framework, which allows us to accurately and efficiently simulate periodic beam lattices. We adopt the multi-lattice formulation of \citet{phlipot_quasicontinuum_2019} but introduce higher-order interpolation in the coarsened regions, while keeping the simplicity of affine interpolation in fully-resolved regions. This allows us to capture the non-uniform deformation modes of bending- and stretching-dominated lattices in 2D and 3D, while only moderately refining the coarsened domain and hence keeping computational expenses low. 

The remainder of this contribution is organized as follows.
We start by introducing the new mixed-order truss QC formulation in Section~\ref{Sec:Method}, before presenting numerical results in Section~\ref{Sec: Numerical Experiments}. After a validation benchmark, we present fracture toughness predictions for 2D and 3D beam lattices, adopting the boundary layer method of \citet{schmidt_ductile_2001}. Section~\ref{sec:Conclusion} concludes our study.

\section{Truss quasicontinuum with mixed interpolation order}
\label{Sec:Method}

\subsection{The quasistatic discrete structural problem}

\begin{figure}[h!]
\centering
\begin{tabular}{ccc}
    \multicolumn{3}{c}{\includegraphics[width=0.95\textwidth]{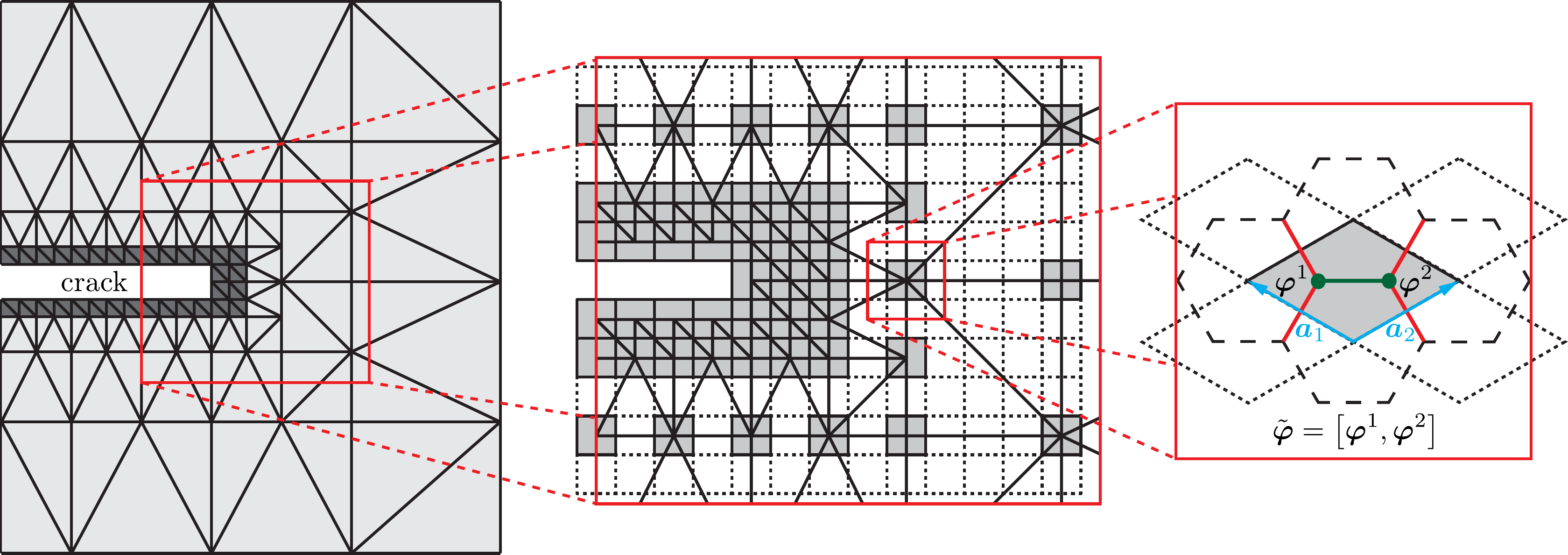}}\\
\end{tabular}
\begin{tabular}{ccc}
     \makebox[0.31\textwidth][c]{(a)} & \makebox[0.32\textwidth][c]{(b)} & \makebox[0.25\textwidth][c]{(c)}
\end{tabular}
    \caption{Schematic overview of the truss QC method. (a) A finite element mesh is used to bridge between regions of full resolution (such as around a crack), here in dark gray and coarse-grained regions (in light gray). (b) Coarse-graining is based on representative UCs (RepUCs), shown in gray, which carry DOFs whose interpolation approximates the deformation of all other UCs (indicated by the dashed lines). (c) A hexagonal unit cell (UC), shown in gray, is spanned by the basis vectors $\bfa_1$ and $\bfa_2$. Nodes within the unit cell and the beams connecting those (i.e., elements of $\calE_u^i$) are colored in green. Beams that connect the UC with neighboring ones (i.e., elements of $\calE^n_u$) are colored in red. The DOFs of both nodes within the UC make up the DOFs $\Tilde{\bfvarphi}$ of the UC.}
    \label{fig:UC}
\end{figure}

We consider a periodic beam lattice $\Omega=\Omega\left(\mathcal{P}, \mathcal{E} \right)$, which is described by a set $\mathcal{P}$ of beam junctions (in the following referred to as \textit{nodes}) and a set $\mathcal{E}$ of beam connections. A beam lattice is constructed by repeating a unit cell (UC) along predefined directions $\mathcal{A}=\left\{\bfa_1,\dots,\bfa_d\right\}$ with basis vectors $\bfa \in \mathbb{R}^d$ in $d$ dimensions. A UC $\Omega_u\subset\Omega$ in the undeformed configuration is characterized by the location $\bfX_{u}$ of its geometric center as well as by its $n_B$ nodes and by the set $\calE_u$ of beams that connect those nodes. Note that we only consider those $n_B$ nodes that uniquely belong to a given unit cell (see the example of a hexagonal UC in Fig.~\ref{fig:UC}c). Here and in the following, we distinguish between beams connecting nodes within the UC (defining the set $\calE_u^\text{i}$) vs.\ those that connect nodes within the UC with nodes in neighboring UCs (defining the set $\calE_u^\text{n}$); see Fig.~\ref{fig:UC}.

\begin{figure}[!t]
    \centering
\begin{tabular}{ccc}
   \includegraphics[width=0.3\textwidth]{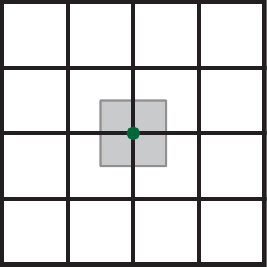}  &\includegraphics[width=0.3\textwidth]{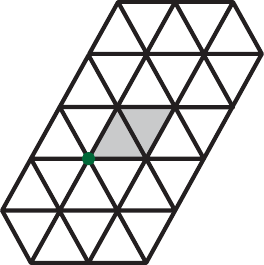}   & \includegraphics[width=0.3\textwidth]{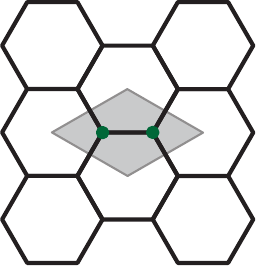} \\
     (a) & (b) & (c)\\
   \includegraphics[width=0.3\textwidth]{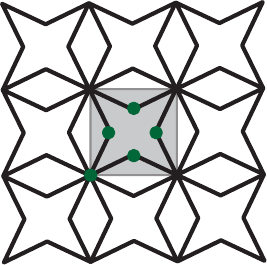} &
   \includegraphics[width=0.3\textwidth]{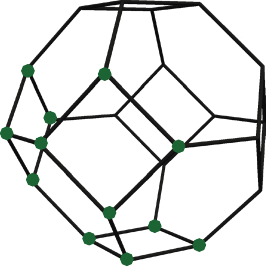}   & \includegraphics[width=0.3\textwidth]{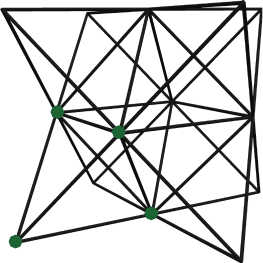} \\
     (d) & (e) & (f) 
\end{tabular}    
    \caption{(a) Square, (b) triangular, (c) hexagonal, (d) star-shaped 2D lattices with their respective UCs highlighted in gray and nodes within the UC shown in green. (e)~Tetrakaidecahedral and (f) octet UCs in 3D, highlighting in green again those nodes that belong to the UC.}
    \label{fig:Multilattice}
\end{figure}

For simple Bravais lattices, e.g., the square (simple cubic) lattice or the triangular lattice in Fig.~\ref{fig:Multilattice}ab, we find $n_B=1$ and $\calE_u = \emptyset$; i.e., each UC contains only a single node, and that node has connections only to neighboring UCs. More complex lattices (e.g., the hexagonal and star-shaped lattices in Fig.~\ref{fig:Multilattice}cd) consist of UCs with $n_B \geq 2$ nodes. These lattices are referred to as \textit{multi-lattices}, adopting the atomistic terminology \citep{ericksen_special_1977}. Each node inside a UC is assumed to have $m$ DOFs. In general, $m$ depends on the chosen model of individual structural members and the dimension $d$. In our case, we consider beam elements, which have $m=3$ DOFs in 2D and $m=6$ DOFs in 3D (including both displacements and rotations). We denote by $\bfvarphi^j\in\Rset^m$ the vector of all DOFs of node $j$. As we combine translations and rotations, we also refer to $\bfvarphi^j$ as the generalized DOFs of node $j$. We further collect all DOFs of a UC $u$, which is centered at $\bfX_u$, in the concatenated vector $\Tilde{\bfvarphi}\left(\bfX_u\right) = \left[ \bfvarphi^1\left(\bfX_u \right), \dots, \bfvarphi^{n_B}\left(\bfX_u \right) \right]$, with $\bfvarphi^j \left(\bfX_u \right)$ being the vector of generalized DOFs of node $j$ within the UC at location $\bfX_u$. Note that our choice of the UC (see Fig.~\ref{fig:Multilattice} for examples) assigns each node in the beam network uniquely to a single UC (and the whole domain is covered by UCs), so that the complete set of (generalized) DOFs of a lattice, $\Tilde\bfvarphi=\{\Tilde{\bfvarphi}\left(\bfX_u\right) : \Omega_u\in\Omega\}$, is the collection of the DOFs of all UCs in the network (where $\bfX_u$ lies in the point lattice spanned by the basis $\calA$).

We here choose an energy-based QC formulation of the problem \citep{eidel_variational_2009,amelang_summation_2015}, which is ideally suited for elastic problems (but also extends to inelastic scenarios with a variational structure \citep{kochmann_quasicontinuum_2016,rokos_variational_2016}).
Assuming a finite-sized beam network and a variational constitutive model, the strain energy of beam $e \in \calE$, which connects nodes $k$ and $l$, is given by $W_e = W(\bfvarphi_k, \bfvarphi_l)$, where $W$ defines the strain energy of a beam (such as the elastic strain energy in case of elastic beam elements). Therefore, the total potential energy of the network is
\begin{align}
    \calI = \sum_{e\in \calE} W_e - \calL, \label{Eq: 03}
\end{align}
with $\calL$ being a linear functional representing the work done by external forces (and/or moments, if rotational DOFs are considered). 
Given the unique decomposition of the model domain~$\Omega$ into non-overlapping UCs $\Omega_u\subset\Omega$, we reformulate the energy in~\eqref{Eq: 03} as
\begin{align}
    \calI  = \sum_{\Omega_u \in \Omega} W_u  - \calL, \qquad\text{where}\quad W_u = \sum_{e\in \calE_u^\text{i}} W_e + \half \sum_{e\in \calE_u^\text{n}} W_e \label{Eq: 04}
\end{align}
is the energy of unit cell $\Omega_u$. Note that $W_u$ depends on $\Tilde{\bfvarphi}(\bfX_u)$ and the DOFs $\Tilde{\bfvarphi}(\bfX_v)$ of all adjacent UCs $v\in\calN(u)$, where $\calN(u)$ defines the neighboring UCs of a UC $u$. The factor $1/2$ in the last term of \eqref{Eq: 04} avoids double-counting the energies of all beams connecting two UCs. Given the above potential energy, the solution of a quasistatic boundary value problem is given by
\begin{align}
    \Tilde{\bfvarphi} = \underset{\Tilde{\bfvarphi} \in V}{\argmin}\ \calI,\label{Eq: 05}
\end{align}
with $V$ being a proper vector space subject to essential boundary conditions.

\subsection{Higher-order quasicontinuum approximation}

Finding a solution to the minimization problem \eqref{Eq: 05} can be prohibitively expensive---even in the simplest case of linear elastic beams. Architected materials can contain extremely large numbers of UCs \cite{shaikeea_toughness_2022}, which leads to prohibitively large numbers of DOFs and hence to prohibitively large systems of equations to solve. Moreover, computing the potential energy in \eqref{Eq: 04} (or its derivatives) requires a prohibitively expensive summation over all UCs in the network. As a remedy, we apply kinematic constraints by introducing an interpolation scheme as well as sampling rules to approximate the potential energy.

\subsubsection{Kinematic constraints}

Akin to the original QC method introduced by \citet{tadmor_quasicontinuum_1996}, we approximate the DOFs of $\Omega$ by a reduced subset. Following \citet{phlipot_quasicontinuum_2019}, we do not select a subset of nodes but instead choose $n_\text{rep} \ll n_{\text{UC}}$ representative UCs (repUCs), where $n_{\text{UC}}$ denotes the set of all UCs in the lattice. Based on those repUCs, we apply an interpolation scheme for the remaining UCs according to 
\begin{align}
    \Tilde{\bfvarphi}\left(\bfX_u \right) = \sum_{r=1}^{n_\text{rep}} N_r\left(\bfX_u \right) \Tilde{\bfvarphi}_r,
\end{align}
where $N_r(\bfX_u)$ are a set of suitable shape functions evaluated at UC locations $\bfX_u$, and $\Tilde{\bfvarphi}_r$ are the DOFs of the repUCs. This interpolation step reduces the number of DOFs significantly from $n_\text{UC}\cdot m$ to $n_\text{rep}\cdot m$.

\begin{figure}[h!]
    \centering
\begin{tabular}{ccc}
    \multicolumn{3}{c}{\includegraphics[width=0.95\textwidth]{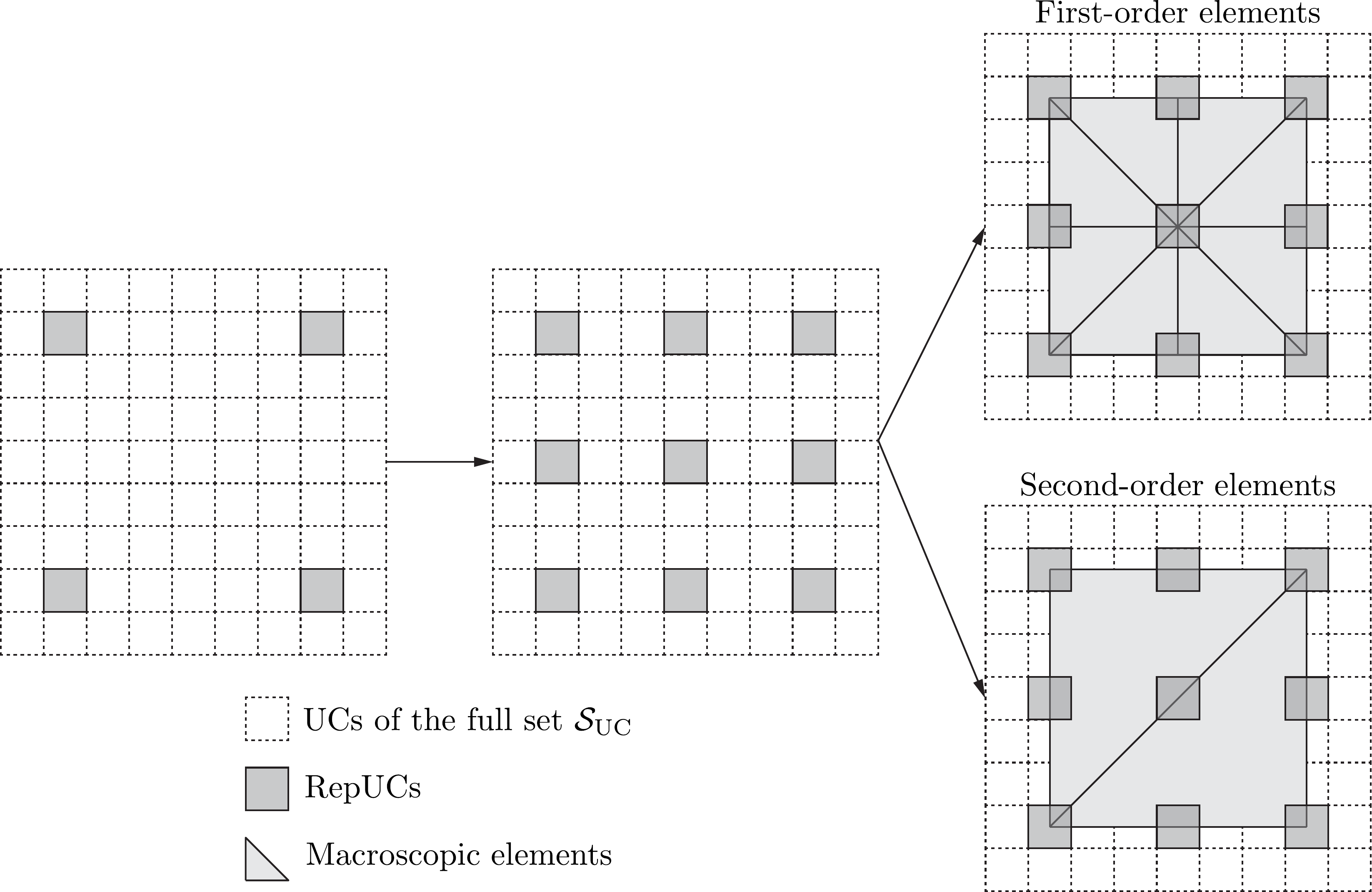}}
\end{tabular}
\begin{tabular}{ccc}
     \makebox[0.1\textwidth][c]{(a)} & \makebox[0.52\textwidth][c]{(b)} & \makebox[0.1\textwidth][c]{(c)}
\end{tabular}    
    \caption{Schematic of the selection of repUCs from the full point set $\calS_{\text{UC}}$ of UCs (dashed lines) and the creation of simplicial meshes. (a) A possible selection of repUCs (gray) with even Bravais coordinates; (b) the addition of mid-edge repUCs from the set $\calS_{\text{mid}}$. (c) Meshing of this set of repUCs leads to either eight first-order or two second-order macroscopic elements, shown in gray.
    }
    \label{fig:RepUC_Selection}
\end{figure}

In the following, we focus on finite element (FE) shape functions, which use the repUC center position $\bfX_u$ to create a conforming Delauney triangulation, so that every node in the resulting FE mesh coincides with a repUC location. We refer to the resulting simplices in the mesh as \textit{macroscopic elements}. (Fig.~\ref{fig:UC}b illustrates a triangulation, whose nodes are repUCs.) When such a simplicial mesh is used, the repUCs can be selected from the full point set of UCs,
\begin{align}
    \calS_{\text{UC}} = \left \{\bfX \in \Omega \, \Big \vert \, \bfX = \sum_{j=1}^d c_j \bfa_j + \bfs_0,\, c_j\in \mathbb{Z}  \right\}, \label{Eq: 07}
\end{align}
with $\bfs_0$ being a global offset vector (the distance from the origin to a UC location) and $c_j$ the \textit{Bravais coordinates}. In the following, we will also use conforming meshes of second-order (quadratic) finite elements. In this case, we first create a simplicial mesh by selecting repUCs from $\calS_\text{UC}$ with even Bravais coordinates $c_j$ (i.e., $c_j/2 \in \mathbb{Z}$), as in Fig.~\ref{fig:RepUC_Selection}a, followed by creating quadratic elements out of the simplicial ones through inserting the mid-edge nodes of all simplicial elements as additional repUC locations (Fig.~\ref{fig:RepUC_Selection}b) from the set
\begin{align}\label{eq:Smid}
    \calS_\text{mid}=\left \{\bfX \in \Omega  \,\Big \vert \,\bfX = \bfX_k + \half \left( \bfX_k-\bfX_l \right),\  \bfX_k,\, \bfX_l \in K_v\right\},
\end{align}
where $K_v$ denotes the vertex locations of macroscopic element $K$ in the simplicial mesh. For linear elements, we turn the edge midpoints into additional vertices, i.e., turn one second-order triangular element in 2D into four first-order ones (Fig.~\ref{fig:RepUC_Selection}c), or one second-order tetrahedral element in 3D into eight first-order ones. Inserting the lattice definition in \eqref{Eq: 07} for $\bfX_k$ and $\bfX_l$ in \eqref{eq:Smid} shows that $\calS_\text{mid}\subset\calS_\text{UC}$; i.e., the RepUCs at edge midpoints according to \eqref{eq:Smid} are automatically valid lattice sites.
This approach allows us to create simplicial meshes (composed of first-order, linearly interpolated constant-strain elements) and second-order meshes (composed of second-order, quadratically interpolated linear-strain elements). In addition, we may combine both element types into meshes that have separate regions of linear and quadratic elements --- this will be a key advantage of our strategy.

\subsubsection{Energy approximation and sampling rules}
\label{Sec:EnergySamplingRule}
In order to reduce the computational expenses of calculating the potential energy in \eqref{Eq: 04}, we introduce, similarly to quadrature rules in the FE method, an approximation to the potential energy as \citep{eidel_variational_2009,amelang_summation_2015}
\begin{align}
    \calI \left(\Tilde{\bfvarphi} \right) \approx \calI^h\left(\Tilde{\bfvarphi} \right) = \sum_{s=1}^{n_s} \omega_s W_s\left(\Tilde{\bfvarphi} \right) - \calL_s \left(\Tilde{\bfvarphi} \right), \label{Eq:10}
\end{align}
with $n_s \ll n_\text{UC}$ sampling sites, in the following referred to as \textit{sampling UCs}, which have weights~$\omega_s$. $\calL_s$ is an approximate (sampled) external force potential. For simplicial meshes, \citet{amelang_summation_2015} derived `\textit{optimal}' sampling rules, whose sampling site locations and weights aimed to minimize the error introduced by \eqref{Eq:10}. (For a specific choice of the weights, their first-order optimal scheme further reduces to the `\textit{central}' scheme proposed by \citet{beex_central_2014}). \citet{phlipot_quasicontinuum_2019} used the first-order optimal sampling scheme for simplicial meshes of repUCs, which we here extend to higher-order elements and the second-order optimal sampling rule.

The choice of sampling UC locations and weights is essential, as it determines the error introduced by \eqref{Eq:10}, which can be detrimental for the accuracy of a coarse-grained simulation. In addition, we must ensure that the chosen sampling scheme recovers the exact, fully-resolved formulation in the limiting case when all UCs are turned into repUCs (i.e., when each and every DOF is being accounted for). This has the benefit that we will not have to explicitly differentiate between fully-resolved and coarse-grained regions and can easily and adaptively refine down to full resolution (as in the so-called `\textit{fully-nonlocal}' QC formulation \citep{amelang_summation_2015}). To this end, we here use the second-order optimal scheme of \citet{amelang_summation_2015} for linear and quadratic elements. 

\begin{figure}[h!]
    \centering
    \centering
\begin{tabular}{cc}
    \includegraphics[width=0.46\textwidth]{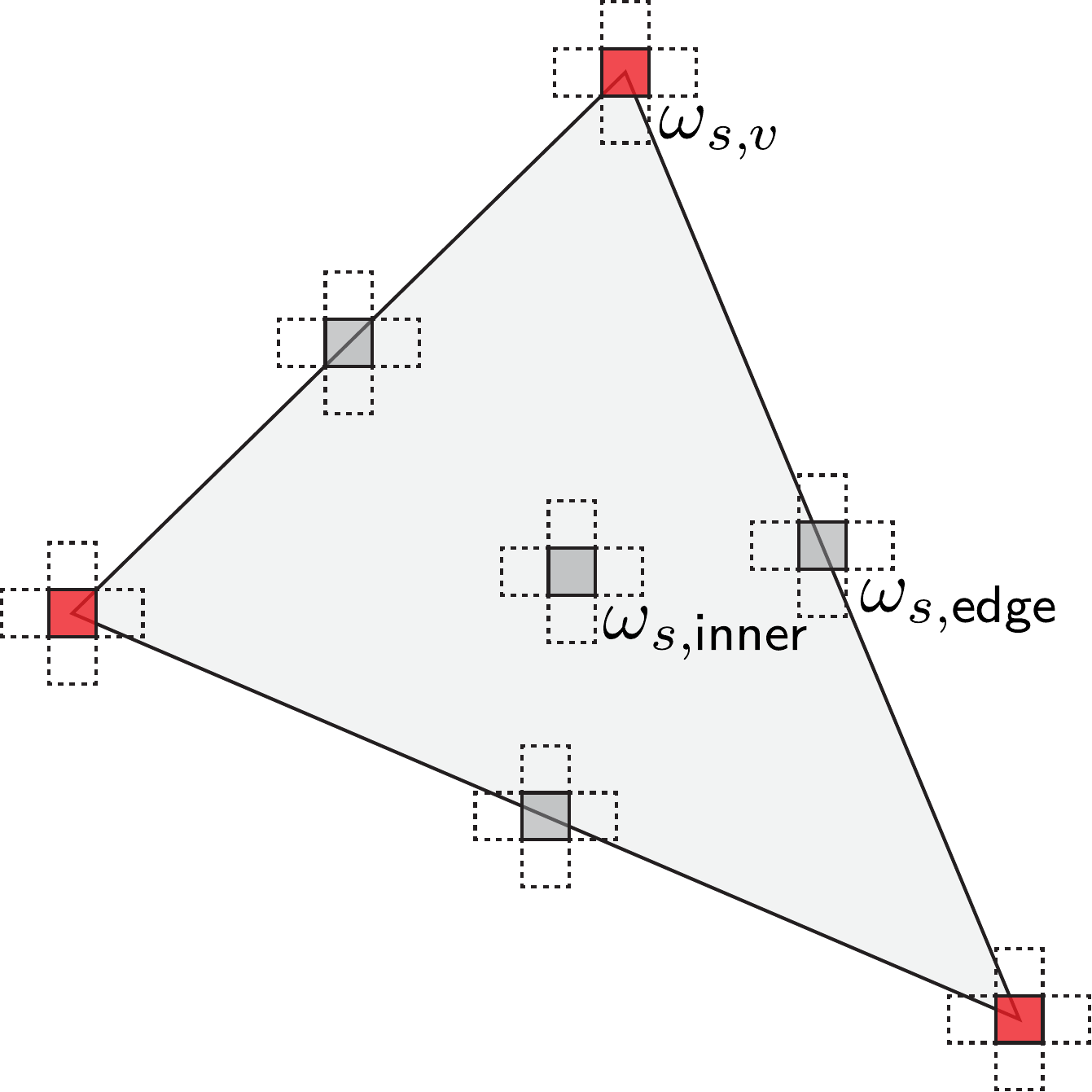} & \includegraphics[width=0.46\textwidth]{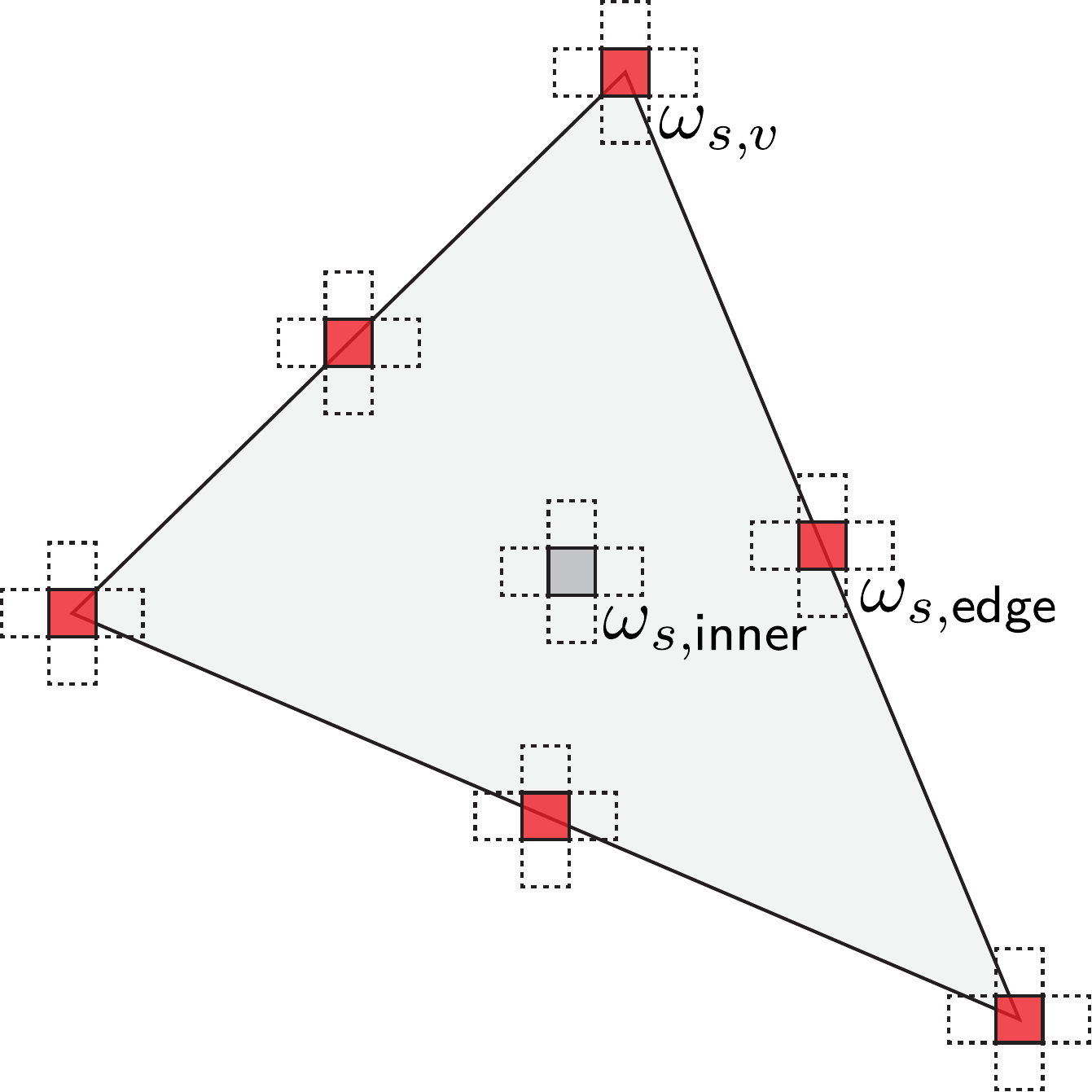}\\
    (a) & (b)\\
    \includegraphics[width=0.46\textwidth]{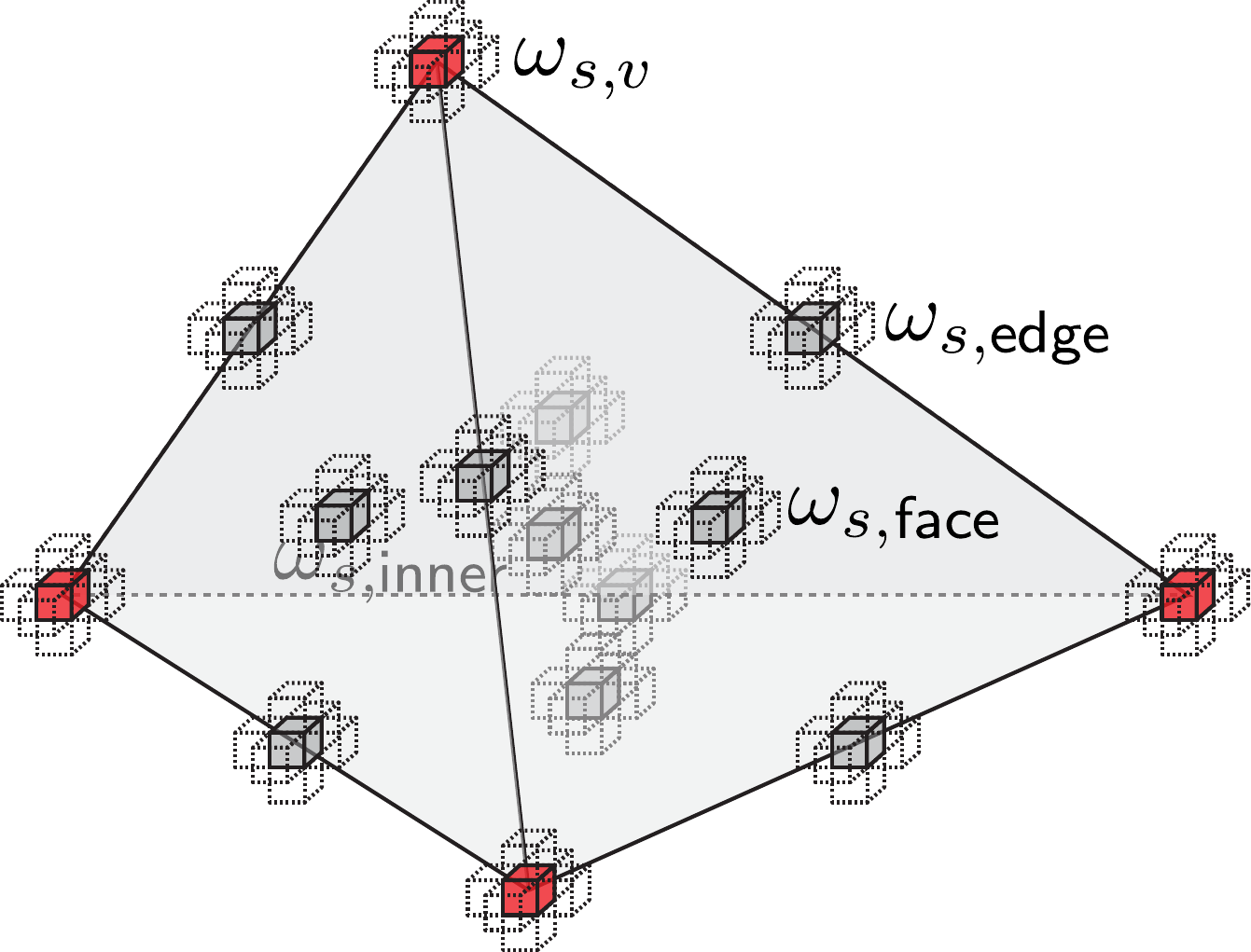} & \includegraphics[width=0.46\textwidth]{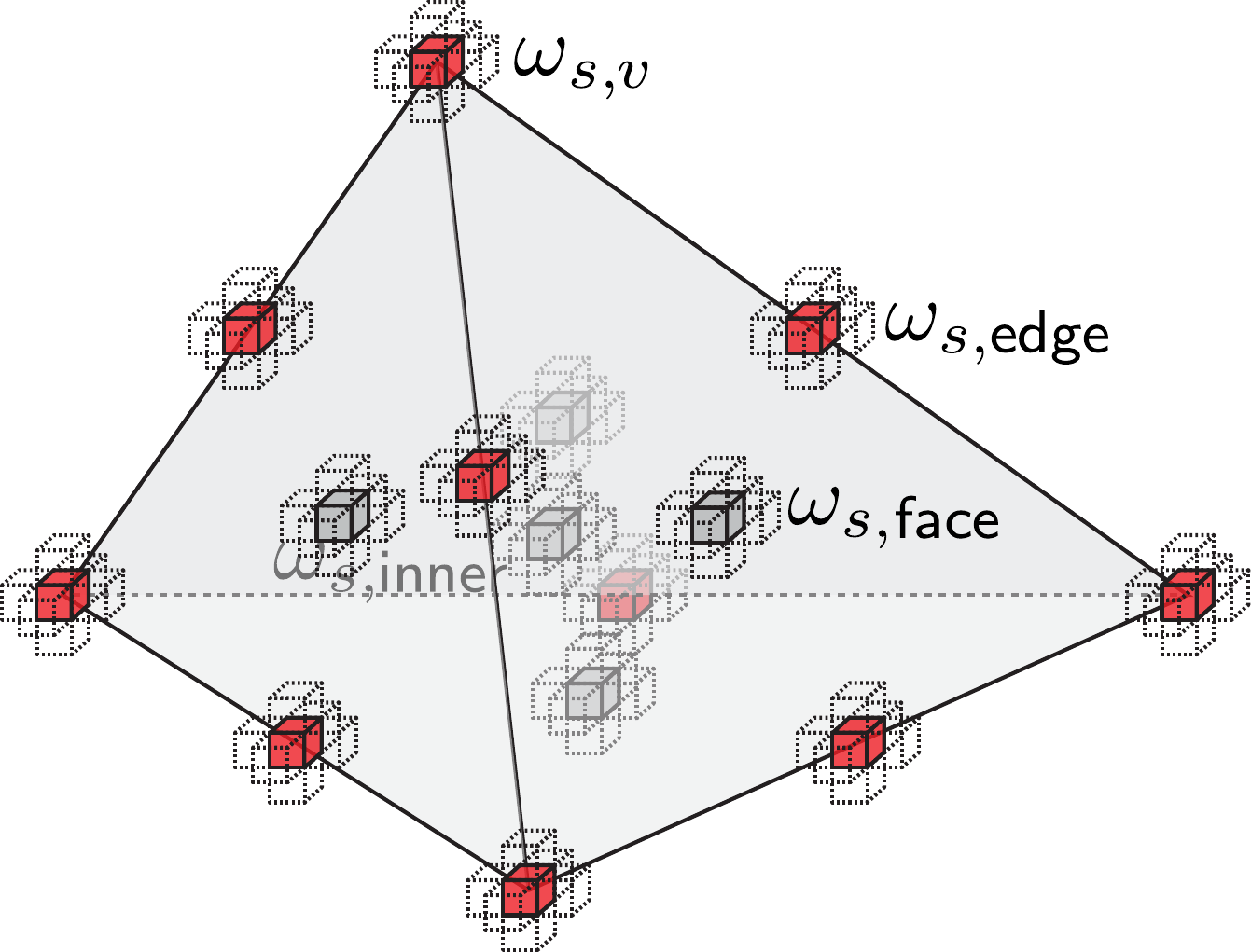}\\
    (c) & (d)    
\end{tabular}
    \caption{Schematic overview of the different elements and the respective choice of sampling UCs for (a) and (c) first-order elements, and (b) and (d) second-order elements in 2D and 3D, respectively. Sampling UCs that are also repUCs are shown in red, additional sampling UCs in gray.
    }
    \label{fig:BaryWeight}
\end{figure}

This scheme is applied to our beam scenario as follows (see also Fig.~\ref{fig:BaryWeight}). In both linear and quadratic elements, the set of sampling UCs includes all vertices and element edge midpoints as well as one sampling UC at the element's barycenter. In 3D, we additionally include one sampling UC per face midpoint. (Note that only for quadratic elements the mid-edge sampling UCs are guaranteed to lie on valid lattice sites, while face midpoints and barycenters do not necessarily lie on valid lattice sites.) For both linear and quadratic elements, this amounts to $3+3+1=7$ sampling UCs per element in 2D, and $4+6+4+1=15$ in 3D (Table~\ref{tab:types}). This defines the locations of all sampling UCs. The energy of each sampling UC is calculated by first computing the DOFs of all nodes within the sampling UC and in its neighboring UCs from the (linear or quadratic) interpolation scheme, based on which the energy follows from \eqref{Eq: 04}. The sampling UC weights are computed as follows.

\begin{figure}[h!]
    \centering
\begin{tabular}{ccc}
   \includegraphics[width=0.3\textwidth]{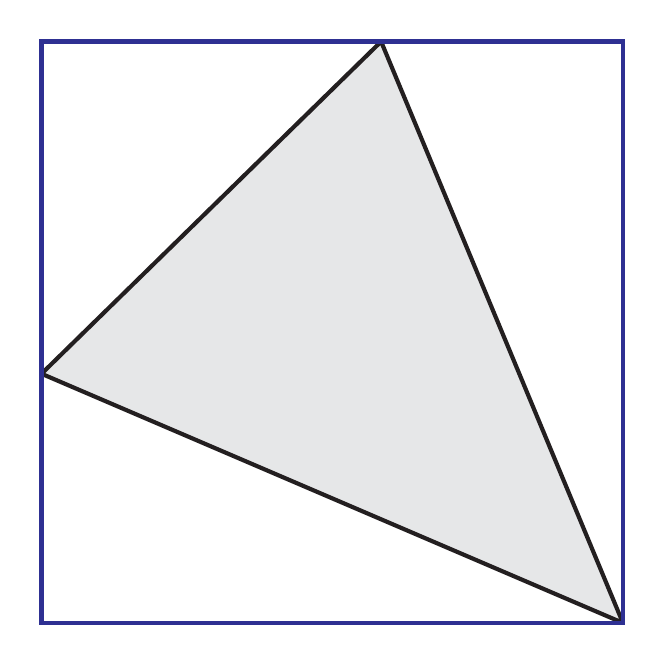}  &\includegraphics[width=0.3\textwidth]{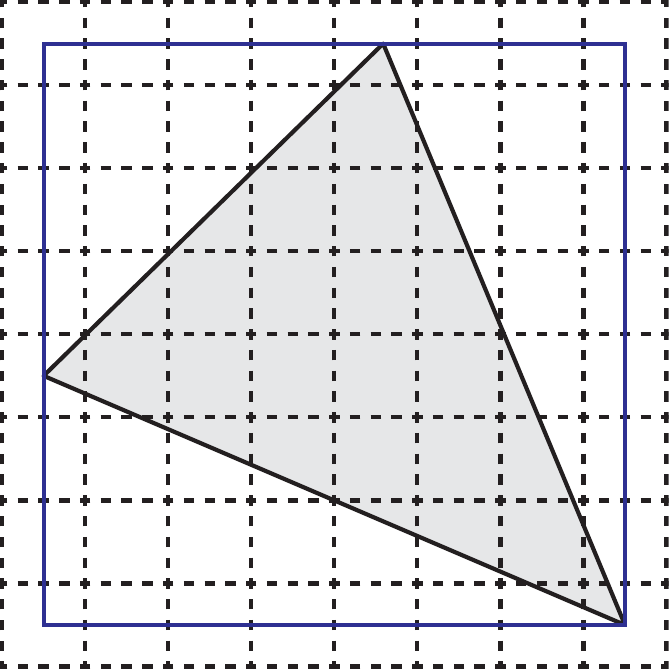}   & \includegraphics[width=0.3\textwidth]{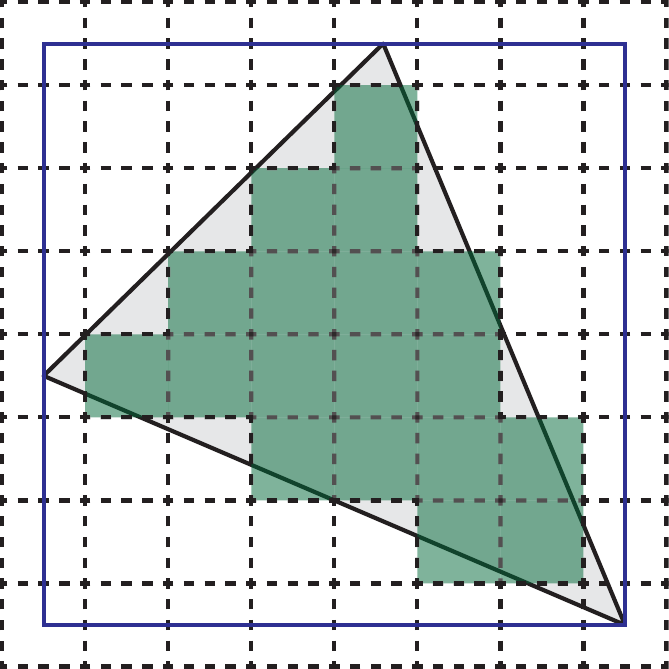} \\
(a) & (b) & (c)
\end{tabular}
    \caption{Schematic showing how the weights of the sampling UCs at a barycenter are calculated in three steps: (a) a macroscopic element (shown in gray) with a box (colored in blue) that fully contains the element; (b) the box is filled with UC locations (shown as dashed lines) as if it was fully resolved; (c) calculation of the distances of each UC to the macroscopic element (UCs with a negative weight are colored in green, others are transparent). UCs on vertices and edges have a distance of zero.}
    \label{fig:BaryWeight2}
\end{figure}

Generally speaking, the sampling UC weights indicate how many UCs inside the element are represented by a given sampling UC in the quadrature rule \eqref{Eq:10}. All sampling UCs on element vertices are given a weight of $\omega_{s,v} =1$ (representing only a single UC). Sampling UCs at edge mid-points are given a weight equal to the number of UCs that lie on a given edge. In 2D, this is calculated by the greatest common divisor (gcd) of the Bravais coordinates of the two endpoints of an edge, i.e.,
\begin{align}
    \omega^\text{(2D)}_{s,\text{edge}} = \text{gcd}\left(Y_2^{(1)} - Y_1^{(1)}, Y_2^{(2)} - Y_1^{(2)} \right) - 1,
    \label{Eq: 11}
\end{align}
with $Y_k^{(j)}$ being the $j^\text{th}$ component of the Bravais coordinate vector (i.e., $c_j$ in \eqref{Eq: 07}) of vertex $k$. In 3D, we analogously define for tetrahedral elements
\begin{align}
        \omega^{\text{(3D)}}_{s,\text{edge}} = \text{gcd}\left(Y_2^{(1)} - Y_1^{(1)}, Y_2^{(2)} - Y_1^{(2)}, Y_2^{(3)} - Y_1^{(3)} \right) - 1.
\end{align}
The weight of the sampling UC at the barycenter is determined by the number of UCs that lie within the (fully-resolved) macroscopic element.
Algorithmically in 2D and 3D, a box is defined that fully contains the macroscopic element (Fig. \ref{fig:BaryWeight2}a). The box is then filled with UC locations of a discrete lattice (Fig.~\ref{fig:BaryWeight2}b), i.e., a subset of $\calS_{\text{UC}}$, by selecting one node of the macroscopic element as the offset vector in \eqref{Eq: 07} and restricting the possible Bravais coordinates to match the size of the box. Lastly, the distances of all those UC locations to the macroscopic element are calculated, so that the weight $\omega_{s,\text{inner}}$ is defined as the number of UCs with a strictly negative distance to the simplex (Fig.~\ref{fig:BaryWeight2}c). UCs on the vertices and edges have a distance of zero and are not counted towards $\omega_{s,\text{inner}}$. The weights $\omega_{s,\text{face}}$ of sampling UCs at face midpoints are computed analogously based on the number of UCs intersecting the respective element faces, which is performed in a numerically efficient manner, as detailed in Appendix~\ref{app:Weights}.

In summary, we use four different types of elements along with the sampling rules listed in Table~\ref{tab:types}. We point out that the chosen sampling rule is $0^\text{th}$-order consistent in up to three dimensions, meaning that it recovers the exact strain energy in 2D and 3D linear and quadratic elements for affine states of deformation. For general loading, the energy is approximated by the chosen sampling rule, the quality of which will be demonstrated by our benchmark examples.

\begin{table}[!h]
    \centering
    \begin{tabular}{c|c|c|c|c|c}
         dimension & element & \multicolumn{4}{c}{number of sampling UCs}\\
         ($d$) & type & vertex & edge & barycenter & face \\
         \hline
         2D & linear/quadratic & $3$ & $3$ & $1$ & $0$\\
         \hline
         3D & linear/quadratic & $4$ & $6$ & $1$ & $4$\\
    \end{tabular}
    \caption{Types of elements and associated sampling rules used in the mixed-order QC formulation. Linear elements are constant-strain triangles and tetrahedra (CST), while quadratic elements are linear-strain triangles and tetrahedra (LST).}
    \label{tab:types}
\end{table}

\subsubsection{Adaptive refinement}
\label{Sec: AdaptiveRefinement}
As in the classical QC context, mesh refinement is a clear strength of the coarse-grained representation---adding resolution where needed (e.g., near the crack tip or in regions of large strains), while efficiently coarse-graining elsewhere. To this end, we implement an adaptive refinement algorithm, which is based on the refinement criterion
\begin{align}
    V(K)^{1/d} f(\bfF_K) > r_0, \label{Eq: 17}
\end{align}
where $V(K)$ is the volume of a macroscopic element $K$ in $d$ dimensions, $f(\bfF_K) = I_2(\bfF_K)$ is chosen as the second invariant of the deformation gradient \citep{knap_analysis_2001,tembhekar_automatic_2017} evaluated at the barycenter of element $K$, and $r_0$ is a a-priori chosen and simulation-specific threshold. Of course, this is only one possible choice for the refinement criterion, whose influence we do not further investigate here (the local violation of centrosymmetry or deformation gradient jumps across elements are possible alternatives). All elements marked for refinement by the above criterion are refined by applying an adapted longest-edge bisection algorithm. Specifically, the restriction of placing nodes of macroscopic elements onto valid lattice sites requires an extra step in the longest-edge bisection algorithm. Therefore, we calculate the position of the bisecting line intersecting the longest edge and choose the valid lattice site closest to it (i.e., RepUCs following \eqref{Eq: 07} with even Bravais coordinates, i.e., $c_j/2 \in \mathbb{Z}$).

\subsubsection{Mixed-order quasicontinuum}
\label{Sec:Mixed-Order}

An essential ingredient of our formulation is its ability to combine first- and second-order elements, which admits the adaptive and seamless transitioning between full resolution and coarse-grained elements with conforming meshes, thus overcoming the deficiencies of prior studies. Almost all prior truss QC methods used linear interpolation, which fails to capture the behavior of bending-dominated topologies \citep{phlipot_quasicontinuum_2019}. By contrast, \citet{beex_quasicontinuum-based_2014} used cubic interpolation of translational DOFs and quadratic interpolation of rotational DOFs, yet this led to different representative sites for translational and rotational DOFs and, therefore, different meshes. Furthermore, their representative sites did not necessarily coincide with lattice sites, and the kinematic constraints led to a non-conforming triangulation at the interface between fully-resolved and coarse-grained domains. 

\begin{figure}[!t]
    \centering
\begin{tabular}{cc}
   \includegraphics[width=0.4\textwidth]{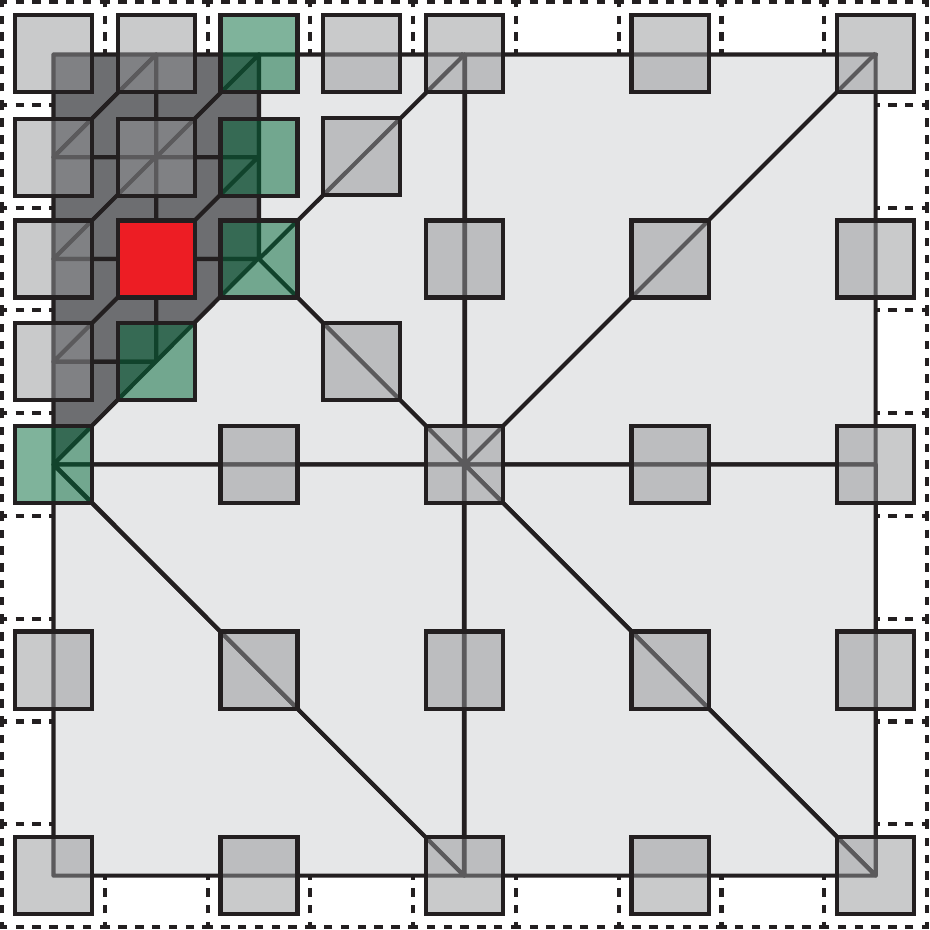}  &\includegraphics[width=0.4\textwidth]{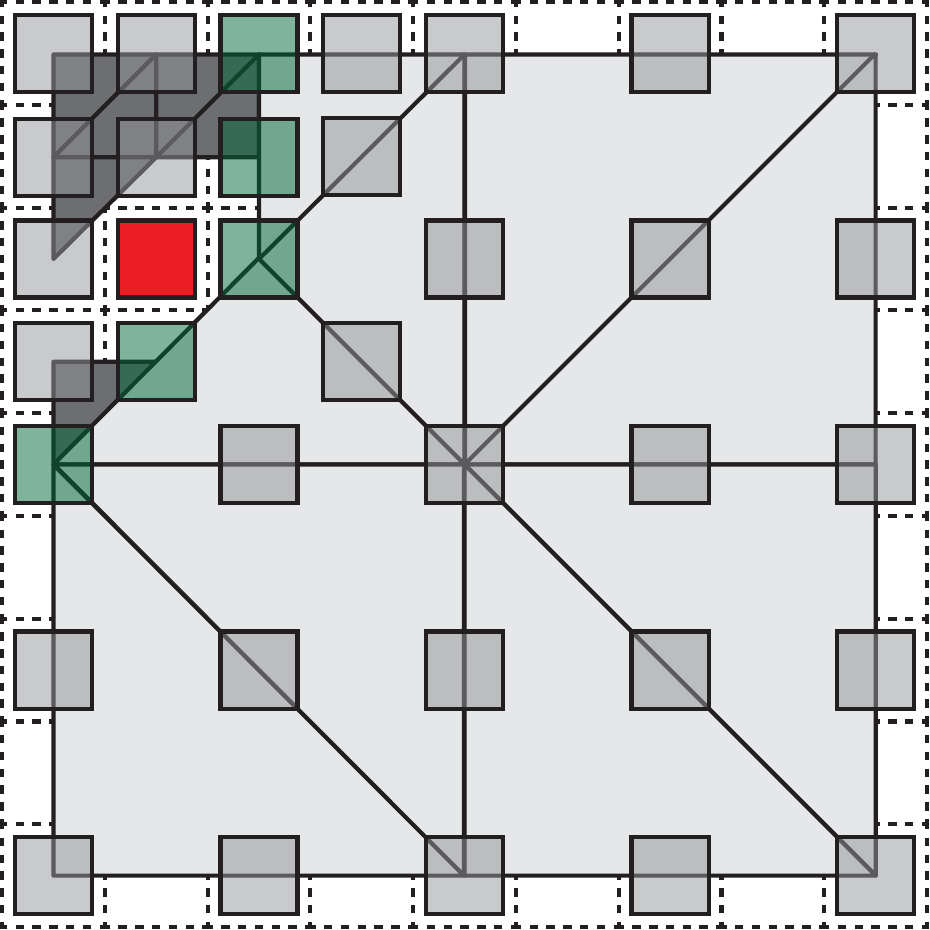}  \\
(a) & (b)
\end{tabular}
\begin{tabular}{cc}
   \includegraphics[width=0.4\textwidth]{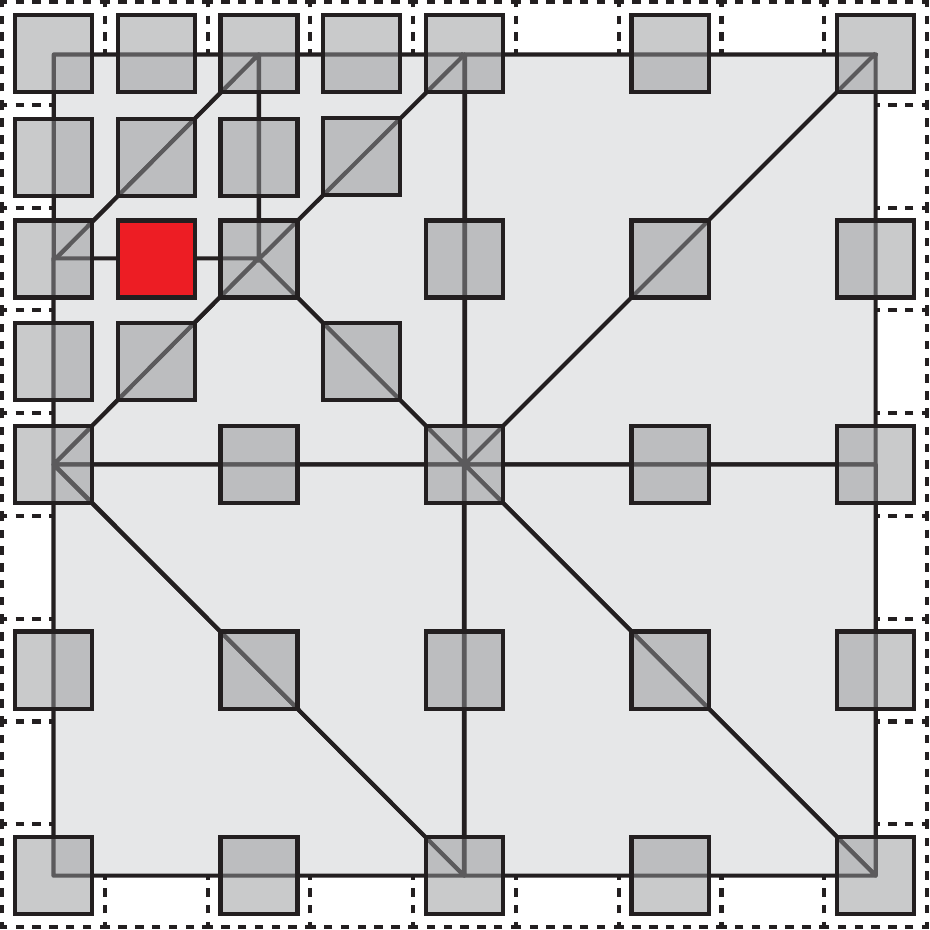}  &\includegraphics[width=0.4\textwidth]{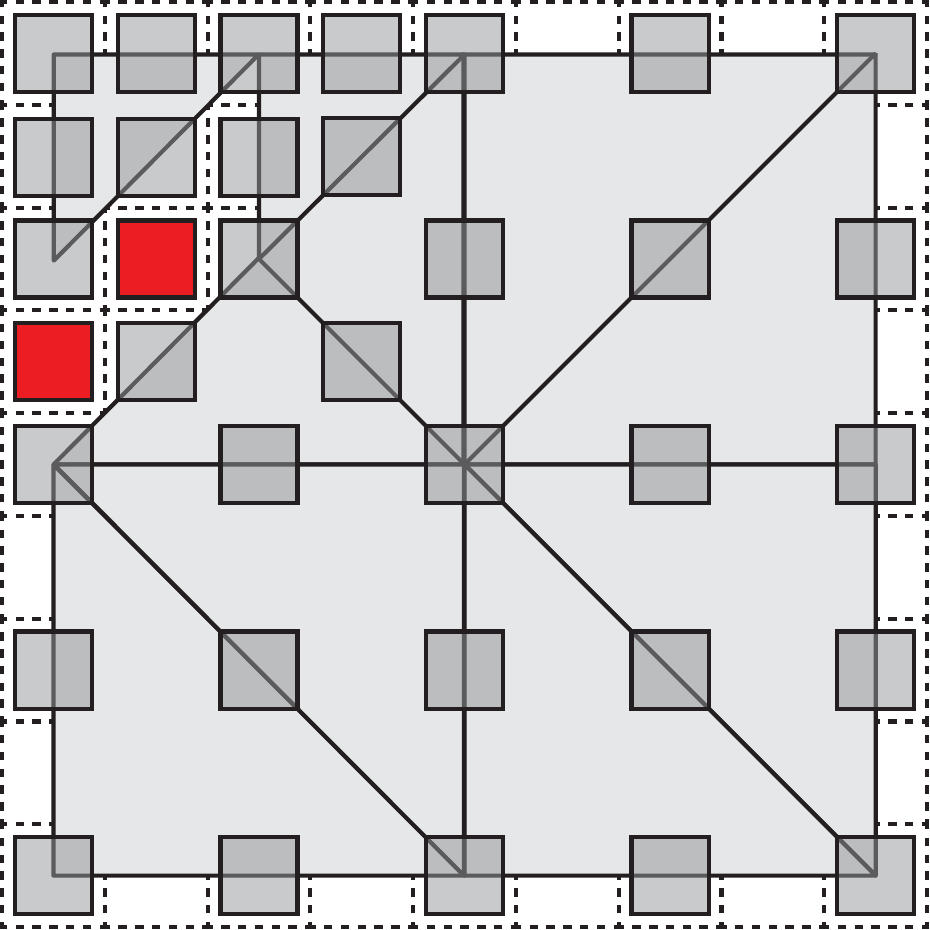}  \\
(c) & (d)
\end{tabular}
    \caption{Demonstrating the benefits of (a,b) a mixed-order mesh over (c,d) a second-order mesh when a UC is to be removed. Light- and dark-gray triangles indicate, respectively, linear and quadratic macroscopic elements. (a) Mixed-order mesh with repUCs in gray and the interface between first-order and second-order elements (the interface in between is shown by green UCs). Removing the red UC in (b) leads to all macroscopic elements being removed that have the removed UC as a vertex -- producing a valid mesh with only the red UC removed. (c) The same scenario with a purely second-order mesh, in which the same UC is to be removed. (d) Removing all macroscopic elements that have the removed UC as a vertex (or mid-edge node) leads to an additional UC being removed, since no valid second-order mesh can be generated with the remaining UCs as nodes.}
    \label{fig:mixed-order}
\end{figure}

Importantly, our fully-nonlocal formulation applies a mesh to the entire simulation domain, including the fully-resolved regions, where it recovers the exact, discrete structural mechanics. To this end, in the fully-resolved region every UC is a repUC and also a sampling UC, as well as a vertex of the simplicial mesh. From a computational perspective, there is little computational overhead for such treatment, while it admits adaptive and flexible remeshing as well as the removal of UCs---which is key to modeling crack propagation. In the fully-resolved domain, we use linear elements, which easily enables the removal of UCs. Fig.~\ref{fig:mixed-order}ab vs.\ Fig.~\ref{fig:mixed-order}cd illustrate, respectively, the removal of a single UC in a fully-resolved linear vs.\ quadratic mesh. While removing a single UC from the linear mesh is trivial, this causes problems for the quadratic mesh and involuntarily leads to removing an additional UC in the proximity of the removed UC (since a quadratic mesh cannot be formed without the removed node). Therefore, we combine linear interpolation (first-order elements) in the fully-resolved domain with quadratic interpolation (higher-order elements) in the coarsen-grained domains. At the interface of the two types of elements, the shape functions are discontinuous in general. However, the lattice's DOFs remain conforming at those interfaces, since we guarantee that every midpoint and vertex of a second-order element is also a vertex of the first-order one. Fig.~\ref{fig:mixed-order}ab show examples of a combined mesh. The fully-resolved region (shaded in dark-gray) consists of linear elements, and it connects compatibly to the adjacent second-order elements (in light-gray) in the coarse-grained domain. At the interface of the fully-resolved region (in green), the mid-edge nodes of the second-order elements coincide with nodes of the first-order elements.

As a further specialty of the fully-nonlocal QC scheme, the energies of sampling UCs (as required in \eqref{Eq:10}) are computed based on the DOFs of their neighboring UCs obtained from exact FE interpolation. (For sampling UCs that are also repUCs this requires the evaluation of shape functions associated with multiple macroscopic elements.)

\subsection{Summary of the proposed truss QC methodology}

As a brief summary, we approximate the mechanics of periodic beam lattices by introducing a fully-nonlocal, mixed-order QC approximation in both 2D and 3D. The simulation domain is discretized into first-order (linear elements) and second-order (quadratic elements). Fully-resolved regions use first-order elements, in which every UC is a repUC as well as a sampling UC, whose weights are chosen such as to recover the exact discrete beam mechanics. Coarse-grained regions use second-order elements, whose nodes are repUCs, while the sampling scheme used for efficient energy, force, and stiffness calculations is based on optimal sampling UC locations and weights. The combination of first- and second-order elements permits a seamless and adaptively evolving transition between the two domains, including the case of UC removal upon mechanical failure. This framework has been implemented within a \CC-based QC library, leveraging the fast solvers of the Portable, Extensible Toolkit for Scientific Computation (PETSc) library \cite{petsc-user-ref, petsc-efficient,PTSCOTCH}, the Toolkit for Advanced Optimization (TAO) \cite{tao-user-ref}, and the MUltifrontal Massively Parallel sparse direct Solver (MUMPS) \cite{MUMPS01, MUMPS02}.

\section{Results}
\label{Sec: Numerical Experiments}

In the following, we report results obtained from applying the above mixed-oder truss QC formulation to a suite of benchmark problems. First, we demonstrate that the presented setup successfully overcomes the previous problem of stretch locking in bending-dominated architectures, followed by simulations of fracture in beam-based architected materials in 2D and 3D.

\subsection{Overcoming stretch locking: Cook's membrane benchmark}

The main problem with previous truss QC formulations based on linear interpolation was a phenomenon, which occurred in bending-dominated lattice topologies: the linear interpolation led to an overly stiff response in the coarse-grained regions. Especially for slender beam networks, bending of beams (whose effective stiffness scales as $k_\text{bend}\propto EI/L^3$) becomes considerably more compliant than stretching (with an effective stiffness $k_\text{stretch}\propto EA/L$). Assuming a rectangular beam cross-section of thickness $t$ and an out-of-plane width of $1$, we arrive at
\begin{equation}
    k_\text{bend}\propto \frac{EI}{L^3} = E \left(\frac{t}{L}\right)^3,
    \qquad
    k_\text{stretch}\propto \frac{EA}{L} = E \frac{t}{L},
\end{equation}
with the base material's Young's modulus $E$, area moment of inertia $I$, cross-section area $A$, and a characteristic beam length $L$. Consequently, in case of slender beams ($t/L\ll 1$) stretching is significantly stiffer than bending. This becomes apparent, e.g., in the hexagonal lattice, which exhibits bending when being stretched, while volumetric expansion or compression is accommodated purely by stretching. Therefore, a slender hexagonal lattice is close to incompressible, which is why a simple linear interpolation in coarse-grained regions leads to problems.
However, an overly stiff response was previously reported \citep{phlipot_quasicontinuum_2019} not only for close-to-incompressible lattices but also for, e.g., the star lattice (whose bulk and Young's moduli are both bending-dominated). This behavior is distinct from volumetric locking and was hence coined \textit{stretch locking} \citep{phlipot_fully-nonlocal_2019}. It was theorized that the linear kinematic constraints in the coarse-grained regions are too restrictive and prevent bending without stretching of individual beams. 

\begin{figure}[b!]
    \centering
\begin{tabular}{ccc}
   \includegraphics[width=0.3\textwidth]{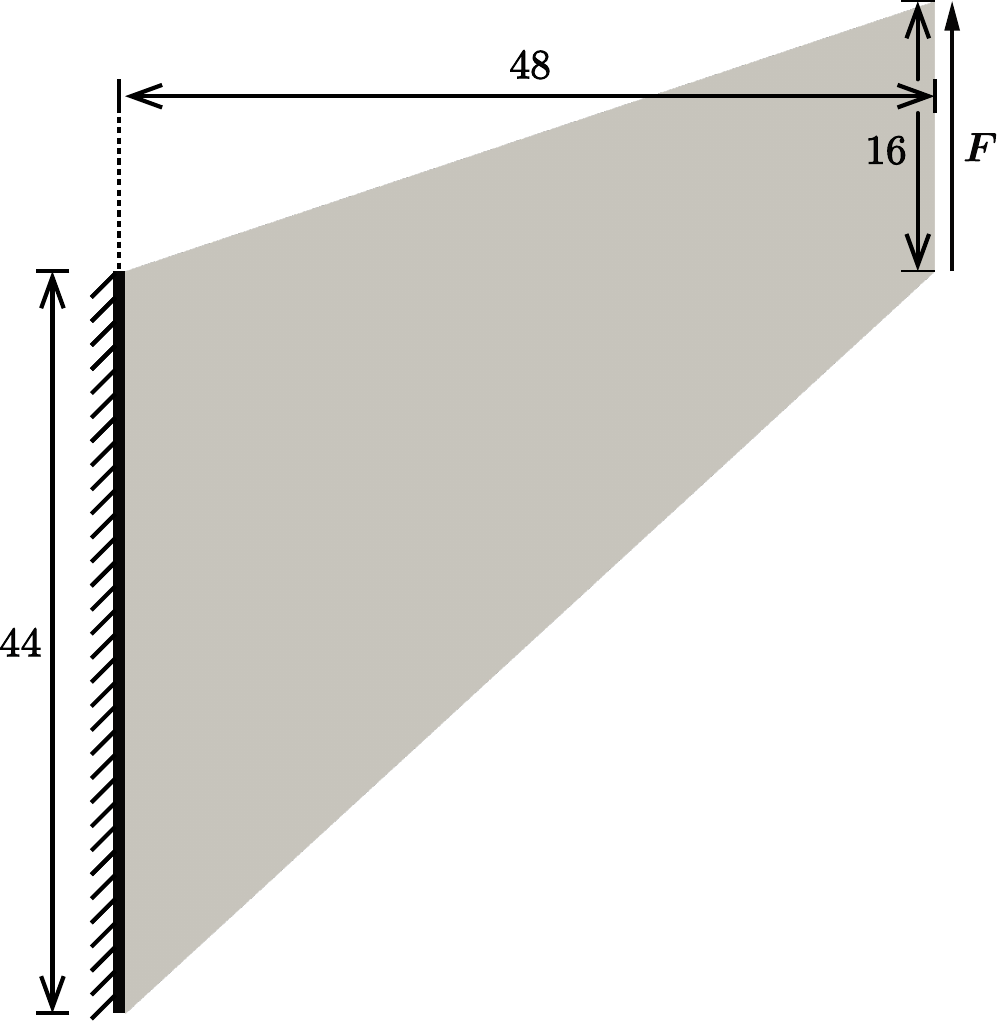}  &\includegraphics[width=0.3\textwidth]{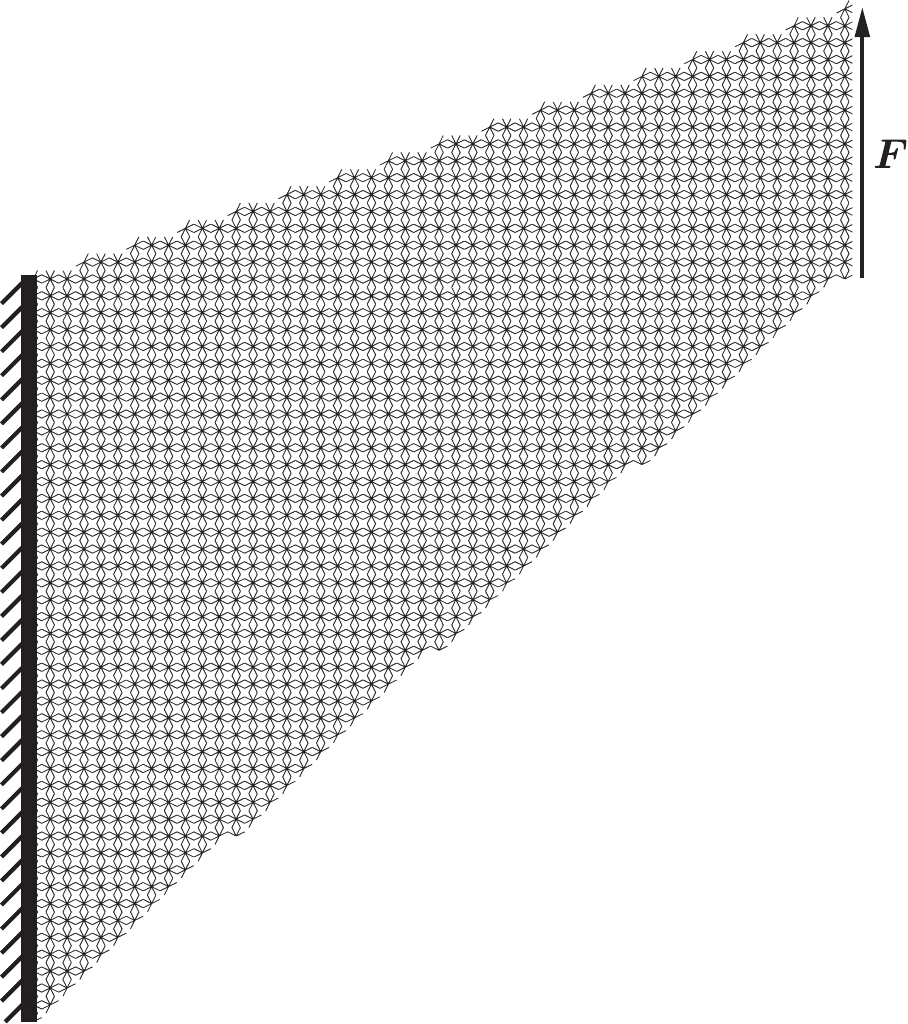}   & \includegraphics[width=0.3\textwidth]{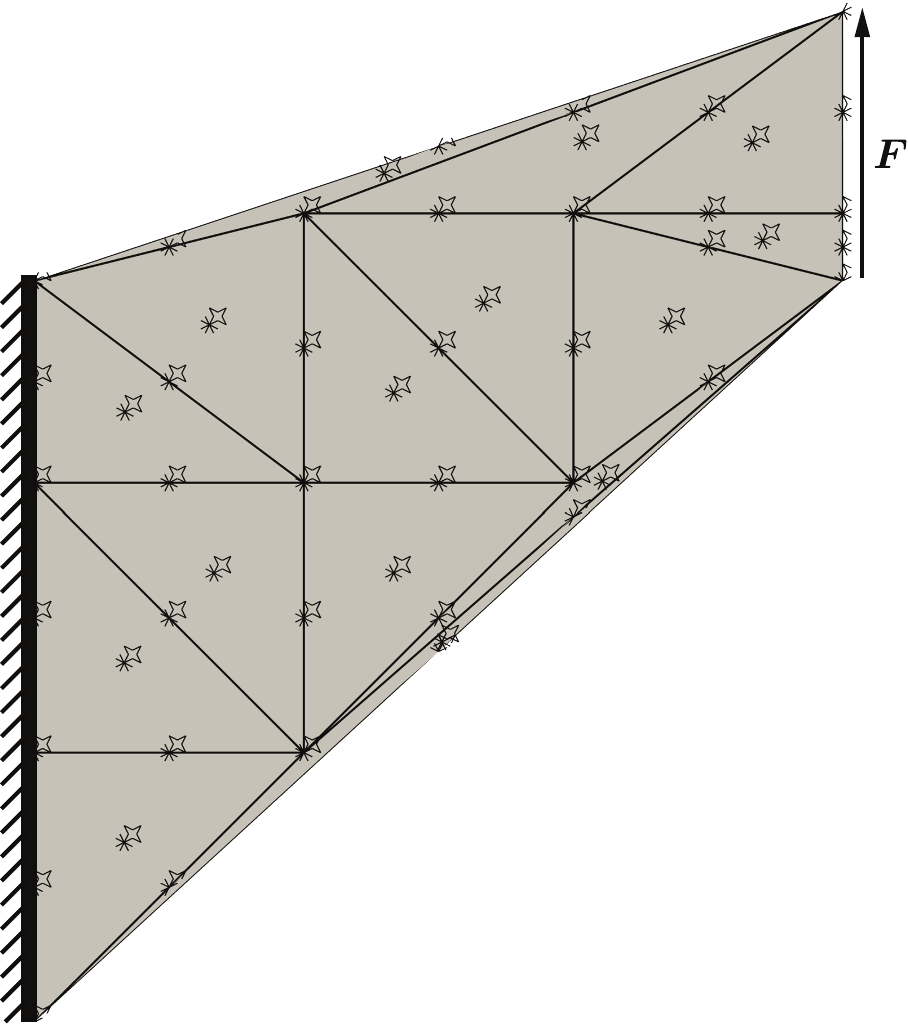} \\
(a) & (b) & (c)
\end{tabular}
    \caption{(a) Cook's membrane geometry in the classical FE setting, (b) the corresponding Cook's membrane problem for a beam lattice with star UCs, and (c) a QC-coarsened representation including all sampling UCs and the corresponding second-order mesh (the repUC density, i.e., the number of repUCs divided by the total number of UCs in the discrete lattice, is about $7.7\%$). In the shown beam lattices, each beam UC has a length of $L=1$. A uniformly distributed force $F$ is applied on the right edge, while the left edge is fixed.}
    \label{fig:Cook_Geometry}
\end{figure}

To confirm this hypothesis and to demonstrate that the present mixed-order formulation does not suffer from the same deficiency, we simulate Cook's membrane, a well-known test case for (volumetric) locking in finite elements \citep{simo_class_1990, cook_concepts_2001}.
Cook's membrane has the 2D geometry shown in Fig.~\ref{fig:Cook_Geometry}, which is held fixed on the left edge, while a uniformly distributed load $F$ is applied on the right edge. Our discrete lattice is constructed by filling the membrane geometry with a predefined truss topology, where each truss member has a specified length $L=1/10$. To apply boundary conditions to the lattice, the left-most nodes in each UC on the left edge are fixed, i.e., the translational and rotational DOFs are set to zero. Similarly, we apply a vertical external force $F$ to the right-most nodes of each UC on the right edge. All beams are modeled as corotational linear elastic Euler-Bernoulli beams \citep{crisfield_consistent_1990}. We have chosen a Young's modulus of $E=430\,$MPa and a Poisson's ratio of $\nu=0.3$ to mimic the trimethylolpropane triacrylate (TMPTA) used by \citet{shaikeea_toughness_2022}.

\begin{figure}[h!]
    \centering
\begin{tabular}{ccc}
    \multicolumn{3}{c}{\includegraphics[width=0.85\textwidth]{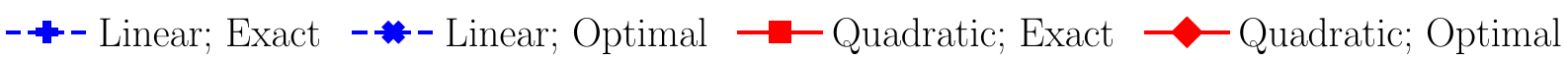}}
\end{tabular}
\centering
\begin{tabular}{cc}
     \includegraphics[width=0.4\textwidth]{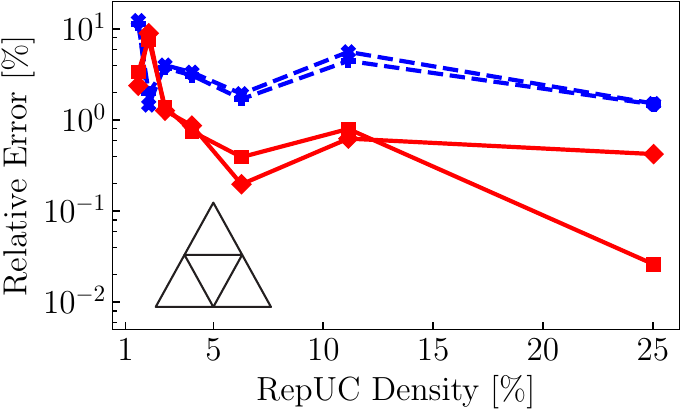} &
     \includegraphics[width=0.4\textwidth]{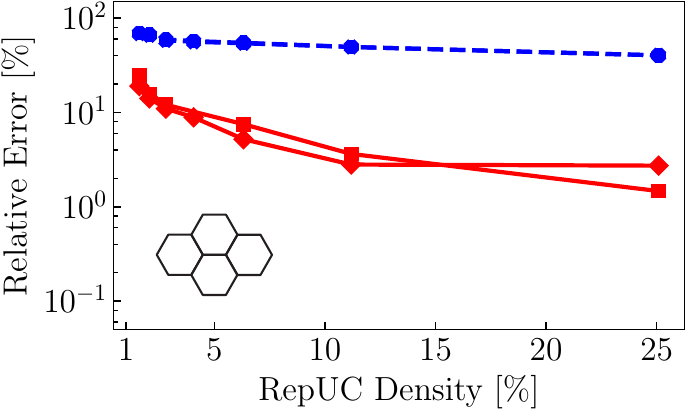}\\
     (a) & (b)
\end{tabular}
\centering
\begin{tabular}{c}
      \includegraphics[width=0.84\textwidth]{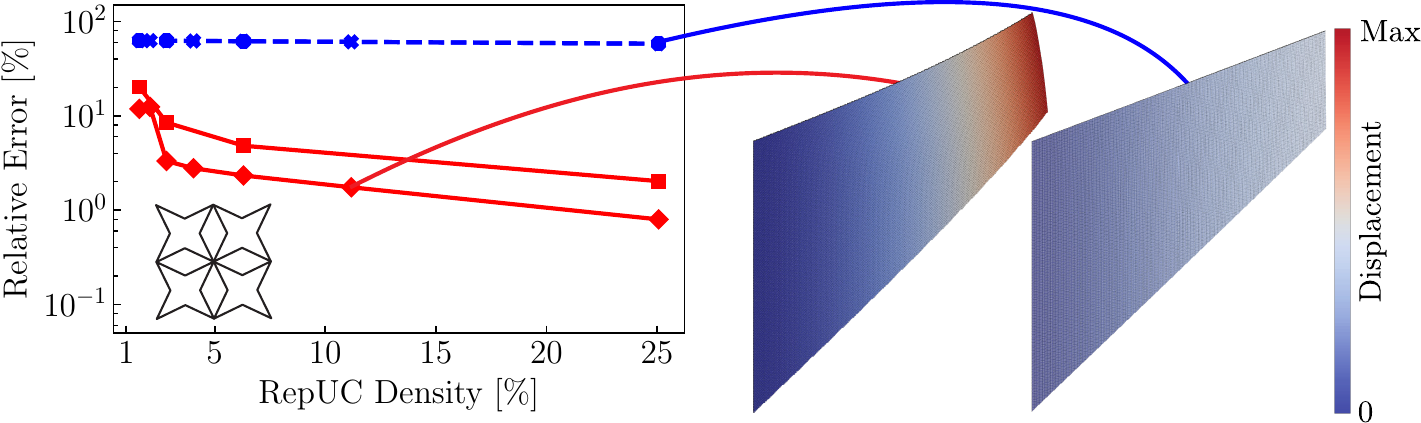}\\
    (c) 
\end{tabular}
\centering
\begin{tabular}{cc}
    \includegraphics[width=0.4\textwidth]{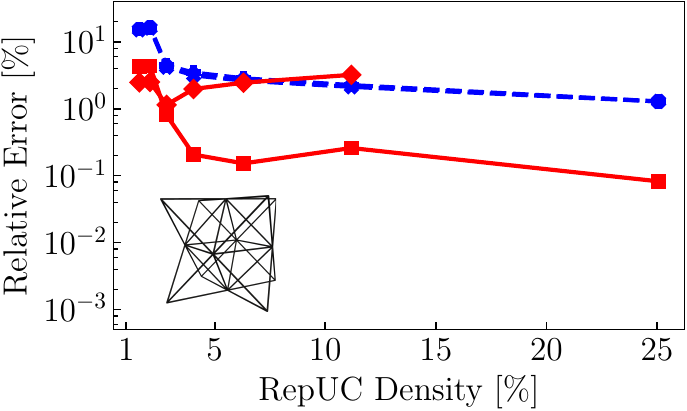}    & 
    \includegraphics[width=0.4\textwidth]{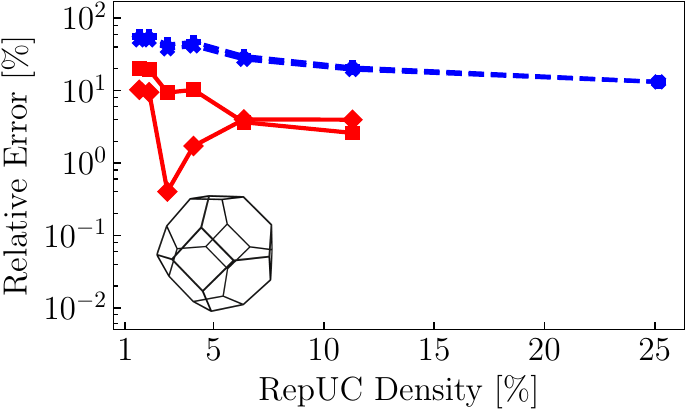}\\
    (d) & (e)
\end{tabular}
\caption{Error of the maximum displacement in the $y$-direction for different UC topologies and different repUC densities in semi-logarithmic plots, including stretching-dominated (a) triangle and (d) octet lattices, and the bending-dominated (b) hexagonal, (c) star and (e) tetrakaidecahedral lattices. Results obtained from second-order elements (red) are compared to those from first-order elements (blue).}
    \label{fig:Cook_Y_Displacement}
\end{figure}

Simulations were performed using the QC method described in Section~\ref{Sec:Method}, and a reference solution was obtained from a fully-resolved discrete beam lattice. Coarsening in the QC model was achieved by selecting every $n^\text{th}$ unit cell as a repUC, where $n\in \left\{2,3,4,6,8\right\}$. The maximum displacement of the lattice in the direction of the applied force $F$ serves as a convenient metric to investigate the locking phenomenon. Fig.~\ref{fig:Cook_Y_Displacement} shows the normalized (with respect to the maximum displacement of the fully-resolved simulation) maximal displacements for the triangular, hexagonal, and star lattices. Each lattice has a relative density of $\Bar{\rho}=1\%$. We further compare two different sampling approaches: first, the second-order optimal sampling rule described in Section~\ref{Sec:EnergySamplingRule} and, second, computing the total energy exactly, i.e., every UC is a sampling UC ($n_s=N_\text{UC}$), and all weights are $\omega_s = 1$ in \eqref{Eq:10}. The level of coarse-graining is indicated by the repUC density, i.e., the number of repUCs divided by the total number of UCs in the discrete lattice.

Results for all investigated lattice topologies are consistent in confirming that the second-order elements perform significantly better than the first-order ones, especially in the regime of low repUC densities (i.e., strong coarse-graining). As conjectured before, errors are considerably higher for bending-dominated topologies than for stretching-dominated ones.
For example, the \textit{triangular} lattice (with a total number of $n_\text{UC}=166{,}431$ UCs) shows, when using first-order interpolation, a relative error of approximately $11.28\%$ for a repUC density of about $1.58\%$, which then drops with increasing repUC density to about $1.69\%$ at a RepUC density of $6.27\%$. With second-order interpolation, the maximal relative error in the same range of repUC densities drops to $3.38\%$ (at $1.58\%$) to $0.39\%$ (at $6.27\%$). The small increase in error around $10$\% is due to the remeshing scheme and not of major concern. Overall, errors are small in this stretching-dominated lattice, with the second-order interpolation achieving lower errors.

By contrast, the \textit{hexagonal} and \textit{star} lattices (which were reported problematic before \citep{phlipot_quasicontinuum_2019}) show significantly larger errors, especially with first-order interpolation. The fully-resolved hexagonal lattice contains $N=55{,}589$ UCs and shows a relative error of about $69.70\%$ at a RepUC density of $1.56\%$, which then decays to $40.29\%$ at a relative density of $25\%$ for the first-order interpolation. With second-order interpolation, errors drop to $15.38\%$ at $2.06\%$ and less than $10\%$ for RepUC densities larger than $6.31\%$, dropping further to $1.47\%$ at a RepUC density of $25\%$. This order-of-magnitude reduction in the error when going from first- to second-order interpolation is repeated for the star lattice (containing $N=144{,}401$ UCs). Errors of $62.79\%$ and $57.99\%$ at repUC densities of, respectively, $1.59\%$ and $25\%$ drop with increasing QC interpolation order to approximately $20.18\%$ and $2.02\%$ at the two above repUC densities. This is an error reduction of about $96\%$. 

Similar results were obtained for 3D lattices, whose Cook's membrane geometries were generated like those in 2D but with an out-of-plane thickness of $3$~UCs. The error for the 3D octet lattice ($N=433{,}203$ UCs) drops from $15.5\%$ to $2.25\%$ at, respectively, repUC densities of $1.59\%$ to $11.18\%$ for the first-order interpolation, and from $4.3\%$ to $0.26\%$ at the aforementioned repUC densities for second-order elements. For the bending-dominated tetrakaidecahedral lattice ($N=69{,}603$), the error reduction from the first-order interpolation to the second-order interpolation is about $87.5\%$ at a repUC density of $11.3\%$.

Therefore, all results in 2D and 3D confirm the prior conjecture that second-order interpolation in the QC coarse-graining overcomes the locking problems of first-order interpolation and yields acceptable errors for both stretching- and bending-dominated topologies. We note that the sampling rule presented in Section~\ref{Sec:EnergySamplingRule} is close to exact energy sampling for first-order elements but deviates significantly from exact sampling in the second-order case, as reported before \citep{beex_quasicontinuum-based_2014}. Nevertheless, for bending-dominated structures the second-order interpolation still considerably outperforms the first-order one in all cases investigated here. Overall, this serves as sufficient confirmation that the second-order elements offers an excellent level of accuracy and overcomes stretch locking, which is why we will use the mixed-order formulation (which combines first-order elements in fully-resolved regions with second-order elements in coarse-grained regions, as outlined in Section~\ref{Sec:Mixed-Order}) in all of the following simulations.

\subsection{Fracture toughness of 2D beam lattices}

\begin{figure}[t!]
\centering
\begin{tabular}{ccc}
    \multicolumn{3}{c}{\includegraphics[width=0.4\textwidth]{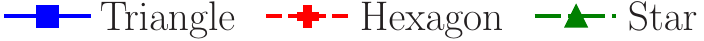}}
\end{tabular}\\
\begin{tabular}{ccc}
    \multicolumn{3}{c}{\hspace{-1.25cm}\includegraphics[width=0.6\textwidth]{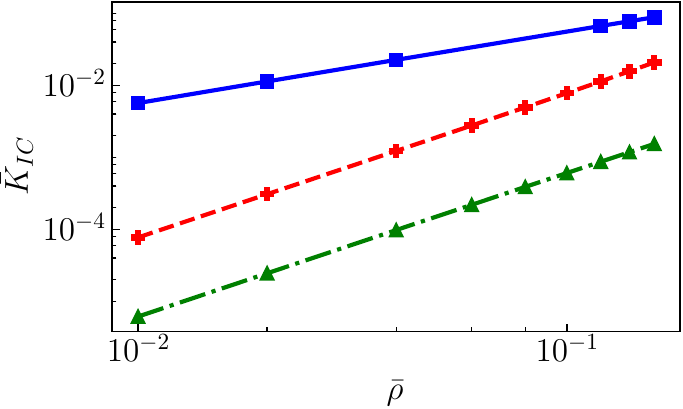}}\\
    \multicolumn{3}{c}{(a)}
    
\end{tabular}
\begin{tabular}{ccc}
   \includegraphics[width=0.22\textwidth]{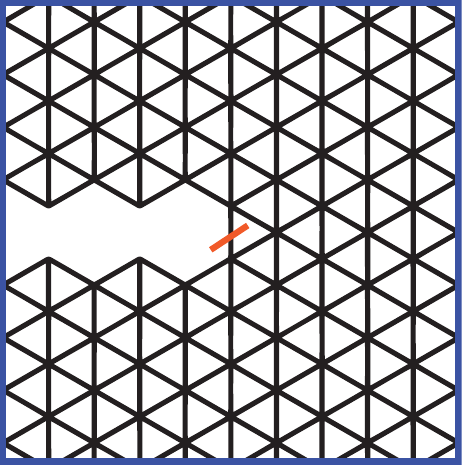}  &\includegraphics[width=0.22\textwidth]{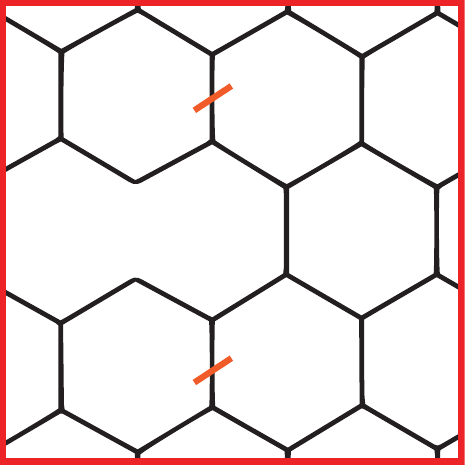}   & \includegraphics[width=0.22\textwidth]{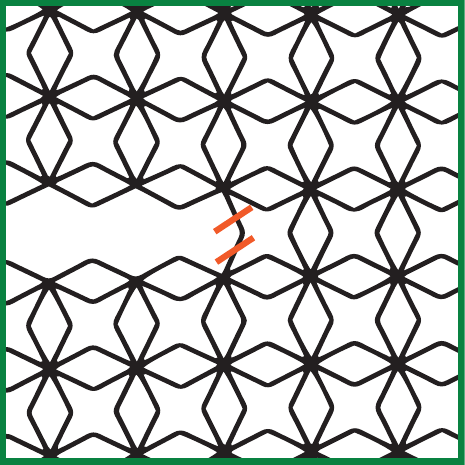} \\
(b) & (c) & (d)
\end{tabular}
\resizebox{\textwidth}{!}{
\begin{tabular}{|c|*{3}{>{\centering\arraybackslash}p{0.1\linewidth}|}*{4}{>{\centering\arraybackslash}p{0.1\linewidth}|}}
\hline
\multicolumn{1}{|c|}{} & \multicolumn{3}{c|}{$D$} & \multicolumn{4}{c|}{$d$}\\
\hhline{|~|-------|}
 & \multicolumn{1}{c|}{Triangle} & \multicolumn{1}{c|}{Hexagon} & \multicolumn{1}{c|}{Star} & \multicolumn{1}{c|}{Triangle} & \multicolumn{1}{c|}{Hexagon} & \multicolumn{1}{c|}{Star} \\
\hline
QC (this work) & 0.5437 & 0.814 & 0.0594 & 0.989 & 2.014 & 1.991  \\
\citet{fleck_damage_2007} & 0.5 & 0.8 & - & 1 & 2 & -  \\
\citet{tankasala_2013_2015} & 0.52 & 0.9 & - & 1 & 2 & -  \\
\hline
\end{tabular}
}
\begin{tabular}{c}
     (e)
\end{tabular}
    \caption{Fracture toughness scaling of different bending- and stretching-dominated lattices. (a) Fracture toughness scaling in a double-logarithmic plot for the stretching-dominated triangular (blue), the bending-dominated hexagonal (red) and star (green) lattices.
    The crack-geometry in the triangle (b), the hexagon (c), and the star (d) lattices, each highlighting the beam that first reaches the failure stress of the parent solid $\sigma_f$ (red).
    (e) Fracture toughness scaling parameters of Eq.~\eqref{Eq: Toughness_Scaling}, based on a linear regression of the data in (a).}
    \label{fig:Toughness_2D}
\end{figure}

The fracture response of beam lattices is a prime example to leverage the new QC formulation. As fracture initiation shows a clearly defined region of interest, viz., the crack flank, we apply the mixed-order QC formulation: first-order elements are initially placed around the crack flank, whereas second-order elements are used in the coarse-grained region as in Fig.~\ref{fig:UC}a. The crack (and hence the crack flank) is defined by removing certain beams from the lattice, depending on the lattice topology. The QC description of the domain makes the removal of beams non-trivial, as those need to be removed from specific UCs. Any beam to be removed is either shared between UCs and hence a member of $\calE_u^\text{n}$ (e.g., in the triangular lattice, Fig.~\ref{fig:Toughness_2D}b), belonging to a single UC and hence a member of $\calE_u^\text{i}$ (e.g., in the hexagonal lattice, Fig.~\ref{fig:Toughness_2D}c), or a combination of both (e.g., in the star lattice, Fig.~\ref{fig:Toughness_2D}d). When removing a shared beam, a second shared beam of a neighboring UC must be removed (following Eq.~\eqref{Eq: 04}). To improve the accuracy, we apply the adaptive refinement scheme of Section~\ref{Sec: AdaptiveRefinement} to the second-order elements in the coarse-grained domain. This process overall leads to QC representations of cracks with full resolution (and first-order elements) around the crack and an efficient coarse-graining (into second-order elemnets) away from the crack. 

\begin{figure}[h!]
    \centering
    \begin{tabular}{c}
    \includegraphics[width=0.65\textwidth]{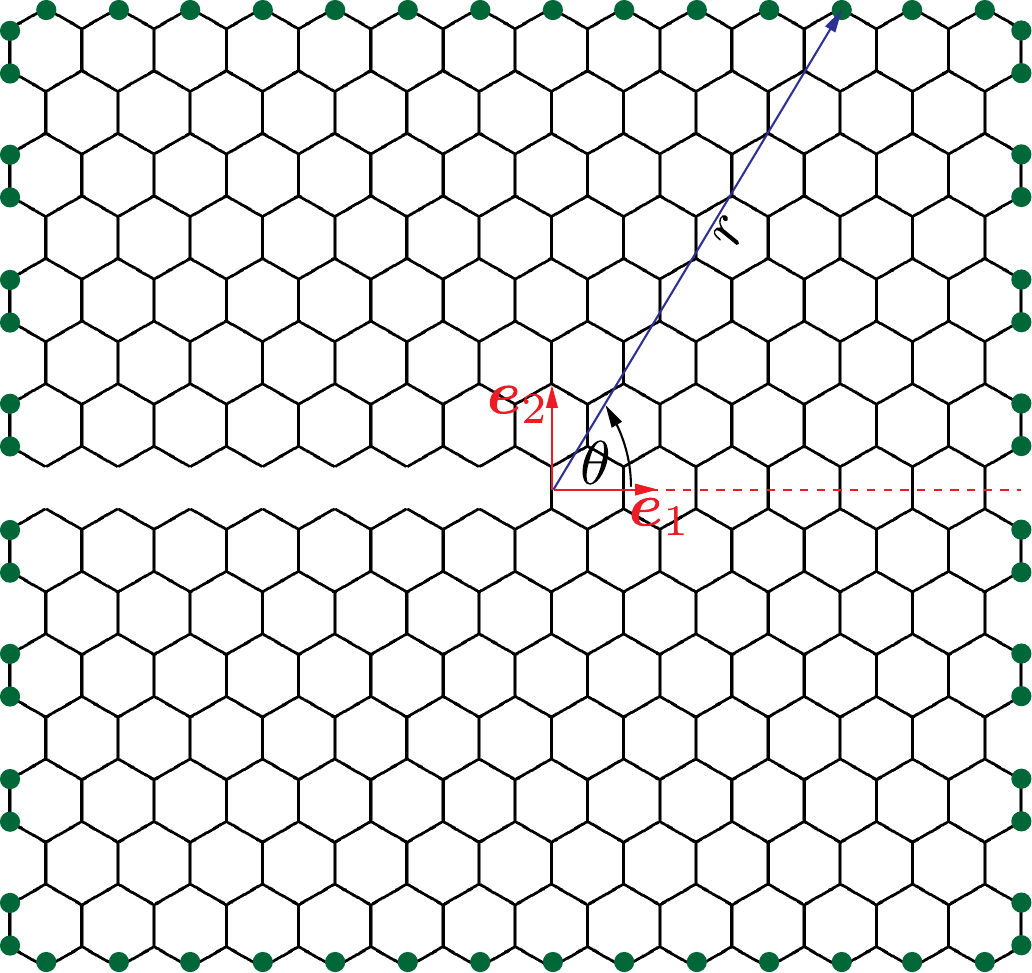}
    \end{tabular}
    \caption{Schematic of the application of the $K$-field to a hexagonal lattice. The coordinate system is placed at the crack tip (red), and Dirichlet boundary conditions $\bfu\left(r,\theta\right)$ are applied to all nodes on the outer boundary (green).}
    \label{fig:K_Field}
\end{figure}

To compute the fracture toughness of 2D beam lattices, we follow the boundary layer method of \citet{schmidt_ductile_2001}, which applies Dirichlet boundary conditions $\bfu\left(r,\theta\right)$, given by the (plane-strain) $K$-field of a crack in an equivalent infinite continuum (which depends on the homogenized elastic stiffness tensor of a given lattice topology) to the outermost nodes of the domain. This approach remedies the finite size of the simulated sample and mimics fracture in an infinite lattice. 
For isotropic lattice topologies (such as the hexagonal and the triangular lattice), we can apply the 2D $K$-fields derived by \citet{williams_stress_1957} (in Cartesian components):
\begin{align}
    \label{eq:Kfield}
    \bfu(r,\theta) = \frac{K_I}{2 \mu_{\text{eff}}} \sqrt{\frac{r}{2\pi}} \left(\frac{3-\nu_{\text{eff,pl}}}{1+\nu_\text{\text{eff,pl}}} - \cos\theta \right) \left[ \cos\left(\frac{\theta}{2}\right)\bfe_1 + \sin\left(\frac{\theta}{2}\right)\bfe_2 \right],
\end{align}
where $K_I$ is the stress intensity factor, $\mu_{\text{eff}}$ the effective (homogenized) shear modulus of the lattice, $r$ and $\theta$ are polar coordinates around the crack tip (Fig.~\ref{fig:K_Field}), and $\nu_\text{eff,pl}$ denotes the effective plane-strain Poisson's ratio of the lattice. Displacements \eqref{eq:Kfield} are imposed as Dirichlet boundary conditions to all nodes on the outer boundary of the simulation domain. By contrast, for anisotropic lattices (such as the star lattice), we apply the formulation of \citet{sih_cracks_1965} for the $K$-field, which is omitted here for brevity. In both cases, we only prescribe translational DOFs, since the impact of additionally prescribing rotational DOFs is negligible \citep{romijn_fracture_2007}.

Brittle fracture is assumed to occur when the maximum tensile stress within an individual beam reaches the critical failure stress $\sigma_f$ of the base material. As the stresses vary in a non-trivial fashion along the length of a beam but the critical points of maximal stresses and beam failure are usually the beam junctions, we calculate the maximum tensile stress $\sigma_t$ in the following based on the axial stretching and bending stresses inside the beam (specifically on the beam surface at the location of maximum stresses), evaluated at its two junctions:
\begin{equation}
    \sigma_t = \max \left(|\sigma_{l}|, |\sigma_{r}| \right),
\end{equation}
with
\begin{equation}
    \sigma_l = \max\left(\sigma_a + \sigma_{b,l}, \sigma_a - \sigma_{b,l}\right)  \quad\text{and}\quad \sigma_r = \max \left(\sigma_a + \sigma_{b,r}, \sigma_a - \sigma_{b,r}\right),
\end{equation}
where $\sigma_a$ is the axial stress, and $\sigma_{b,l}$ and $\sigma_{b,r}$ denotes the maximum bending stresses at the left and right nodes of the beam, respectively.
Although normalization of results makes the exact moduli irrelevant, each beam is modeled with a Young's modulus of $E=430\,$MPa, a Poisson's ratio of $\nu=0.3$, a critical stress of $\sigma_f=11\,$MPa, and a shear correction factor of $1.2$, to again mimic the TMPTA samples of \cite{shaikeea_toughness_2022}.

\begin{figure}[b!]
    \centering
    \begin{tabular}{ccc}
    \multicolumn{3}{c}{\includegraphics[width=0.4\textwidth]{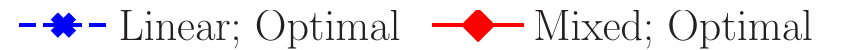}}
    \end{tabular}
\begin{tabular}{ccc}
\includegraphics[width=0.3\textwidth]{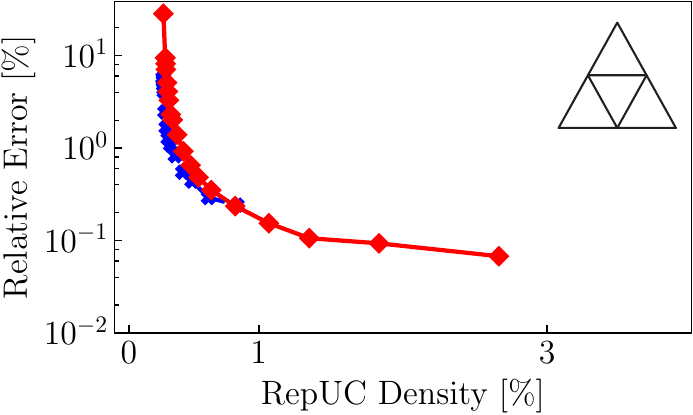} & \includegraphics[width=0.3\textwidth]{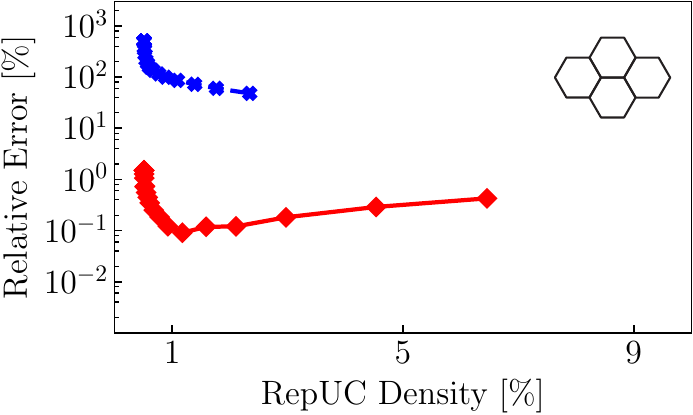} & \includegraphics[width=0.3\textwidth]{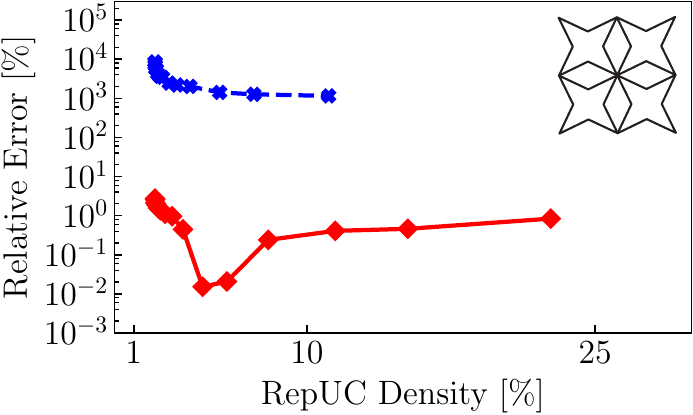}\\
(a) & (b) & (c)
\end{tabular}
    \caption{Relative error in the strain energy of the beam lattice vs.\ repUC density in a semi-logarithmic plot for (a) the triangular, (b) the hexagonal, and (c) the star lattice. Each consecutive data point is obtained by adaptively refining the initial mesh; i.e., starting from an initial threshold value of $r_0$ (at the lowest repUC density), reducing the threshold by $20\%$ and minimizing \eqref{Eq:10} after each refinement iteration. The first-order interpolation (blue) is compared with the mixed-order interpolation (red), both using optimal sampling.}
    \label{fig:Strain_Energy_2D}
\end{figure}

As we now leverage the full mixed-order QC setup with adaptive remeshing, results will depend on the level of mesh refinement, which in turn depends on the chosen threshold for adaptive refinement, i.e., $r_0$ in~\eqref{Eq: 17}. Each initial mesh has full resolution only on the crank flank and hence requires refinement to accurately capture the stress distribution in the beams. To this end, we refine the mesh gradually by sequentially reducing $r_0$ in increments of $20\%$ and minimizing \eqref{Eq:10} after every refinement step, whose results are shown in Fig.~\ref{fig:Strain_Energy_2D}. Starting with the lowest repUC density, each data point corresponds to the results of a successive refinement step based on the incrementally reduced $r_0$-value, showing a reduction in the error with increasing refinement.
All results were obtained from the optimal sampling rule, as it has proven accurate in the previous section and because exact energy sampling is too costly given the amount of UCs in the following examples. The fully-resolved triangular lattice consists of $N=4{,}196{,}820$ UCs with approximately $12{,}590{,}460$ DOFs, while the fully-resolved hexagonal and star lattices consist of, respectively, $N=3{,}672{,}831$ and $N=804{,}609$ UCs, leading to approximately $7{,}345{,}662$ and $12{,}069{,}135$ DOFs, respectively. Here and in the following, we chose an initial threshold value of $r_0 = 0.012$ for the hexagonal lattice and $r_0 = 0.04$ for the star lattice. For both bending-dominated topologies we chose $K_I=10^{-6}$. As the stretching-dominated triangular lattice is significantly stiffer, we increased the applied value of $K_I$ to $K_I =6.27\cdot10^{-2}$ and used the initial refinement threshold $r_0=0.3$.

Results show that the strain energy of the (stretching-dominated) triangular lattice accurately is captured for both linear and mixed-order interpolation, as may be expected, with errors below $1\%$ after $8$ refinement steps (which corresponds to repUC densities below $0.5\%$ or a value of $r_0 \approx 0.05$). By contrast, the first-order interpolation massively overpredicts the strain energy of the hexagonal and star lattices, while the mixed-order formulation achieves errors as low as $0.091\%$ for the hexagonal lattice (after $14$ refinements or a repUC density of about $1.17\%$) and $0.015\%$ in the star lattice (after $9$ refinements). Those minimal errors correspond to refinement thresholds of, respectively, $r_0 \approx 0.0017$ and $r_0 \approx 0.0054$. The strategy in our simulations will hence be to pick the aforementioned initial values of $r_0$ and to perform adaptive refinement for a reasonable number of steps.

\begin{figure}[b!]
    \centering
    \begin{tabular}{ccc}
    \multicolumn{3}{c}{\includegraphics[width=0.4\textwidth]{Legend_Fracture_LowRes.pdf}}
    \end{tabular}
\begin{tabular}{ccc}
   \multicolumn{3}{c}{\includegraphics[width=0.9\textwidth]{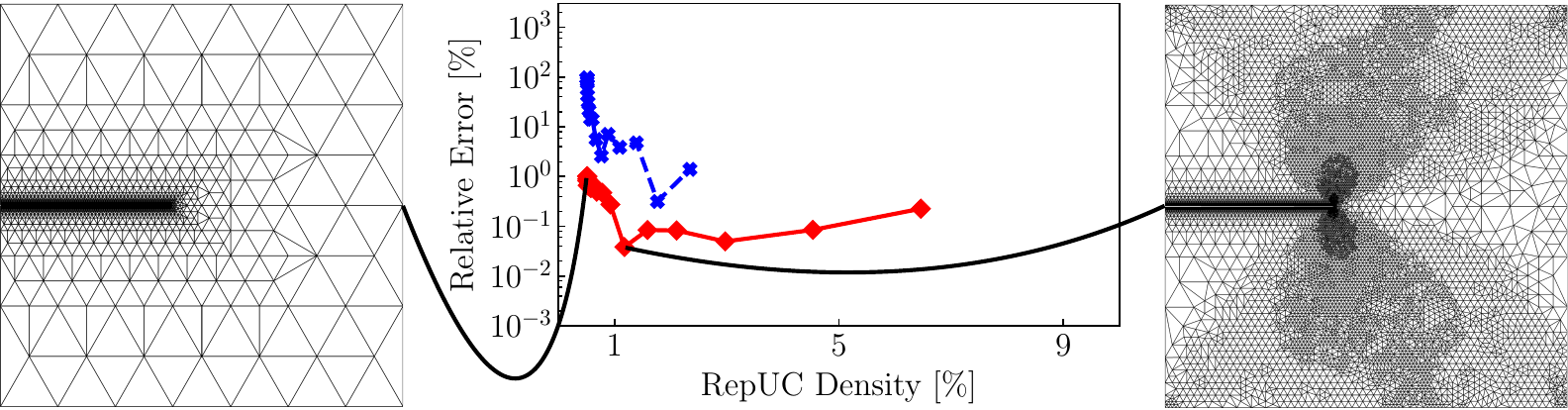}}
\end{tabular}
\begin{tabular}{ccc}
    \makebox[0.0\textwidth][c]{(a)} & \makebox[0.6\textwidth][c]{(b)} & \makebox[0.00\textwidth][c]{(c)}
\end{tabular}
\begin{tabular}{ccc}
\multicolumn{3}{c}{\includegraphics[width=0.9\textwidth]{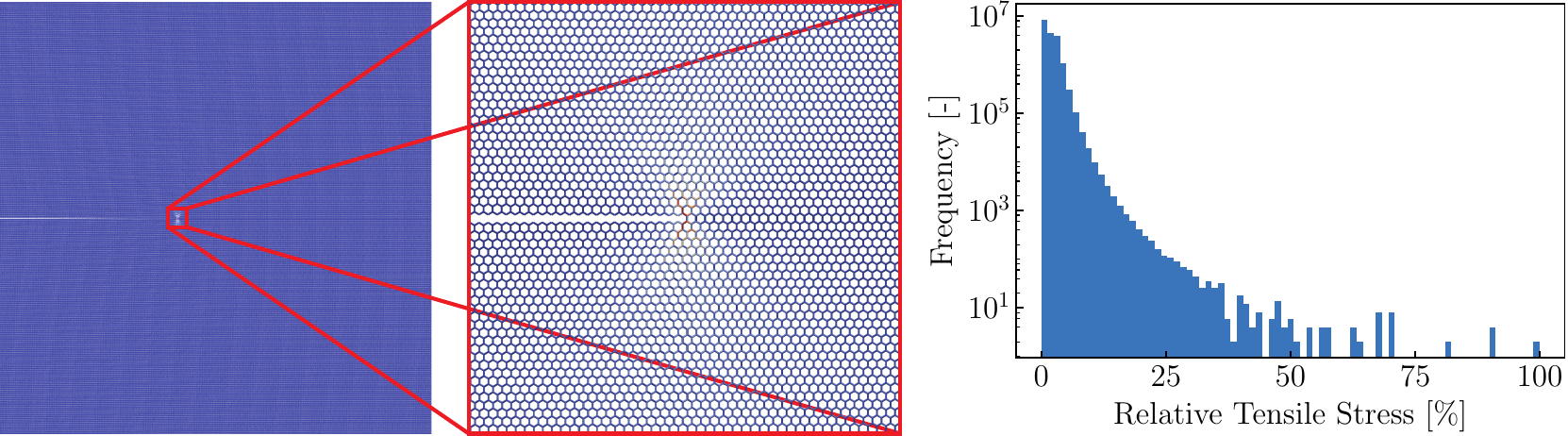}}
\end{tabular}
\begin{tabular}{ccc}
     \multicolumn{3}{c}{\makebox[0.25\textwidth][c]{(d)} \makebox[0.3\textwidth][c]{(e)} \makebox[0.35\textwidth][c]{(f)}}
\end{tabular}
\caption{Results for mode-I loading of a pre-notched \textit{hexagonal} lattice. (a) The initial mixed-order QC mesh. (b) Convergence of the maximum tensile stress $\sigma_t$ with adaptive refinement for $20$ refinement steps, comparing first- (red) and mixed-order elements (blue). (c) Refined mixed-order mesh after $14$ adaptive refinement steps (i.e., the number of refinement steps corresponding to the minimal error in the strain energy). (d) Distribution of the tensile stresses in the entire simulation domain and (e) in the magnified region around the crack tip. (f) A histogram using $80$ bins, showing the relative (normalized by the maximum tensile within the hexagonal lattice) tensile stresses and the frequencies in which they appear in the complete model.}
    \label{fig:Convergence_Fracture_Hexagon}
\end{figure}

\begin{figure}[t!]
    \centering
    \begin{tabular}{ccc}
    \multicolumn{3}{c}{\includegraphics[width=0.4\textwidth]{Legend_Fracture_LowRes.pdf}}
    \end{tabular}
\begin{tabular}{ccc}
   \multicolumn{3}{c}{\includegraphics[width=0.9\textwidth]{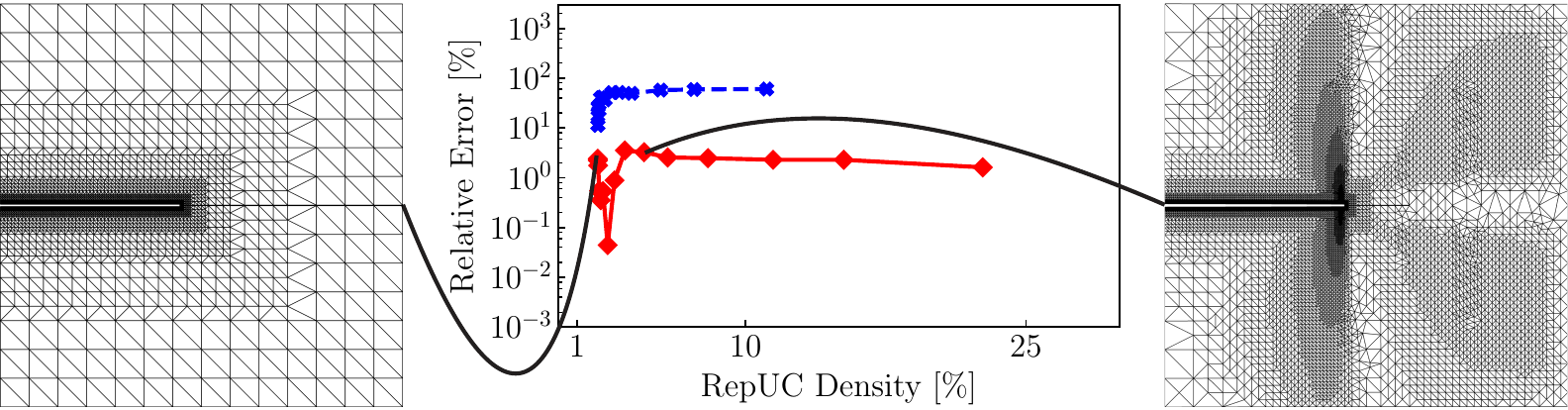}}

\end{tabular}
\begin{tabular}{ccc}
    \makebox[0.0\textwidth][c]{(a)} & \makebox[0.6\textwidth][c]{(b)} & \makebox[0.00\textwidth][c]{(c)}
\end{tabular}
\begin{tabular}{ccc}
\multicolumn{3}{c}{\includegraphics[width=0.9\textwidth]{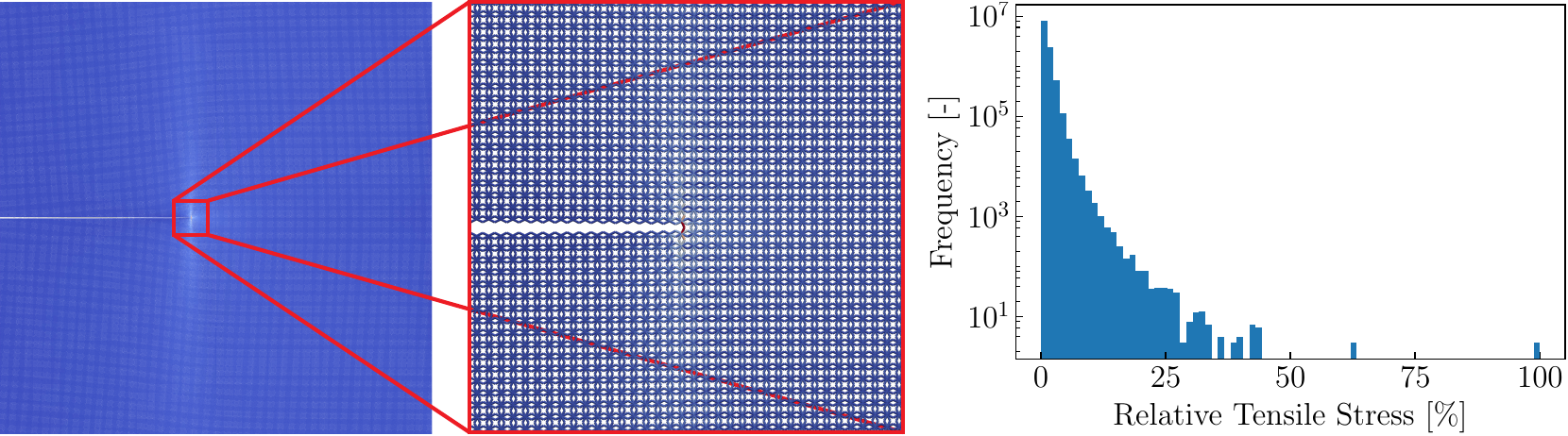}}
\end{tabular}
\begin{tabular}{ccc}
\multicolumn{3}{c}{\makebox[0.25\textwidth][c]{(d)} \makebox[0.3\textwidth][c]{(e)} \makebox[0.35\textwidth][c]{(f)}}
\end{tabular}
\caption{Results for mode-I loading of a pre-notched \textit{star} lattice. (a) The initial mixed-order QC mesh. (b) Convergence of the maximum tensile stress $\sigma_t$ with adaptive refinement for $15$ refinement steps, comparing first- (red) and second-order elements (blue). (c) Refined mixed-order mesh after $9$ adaptive refinement steps (i.e., the number of refinement steps corresponding to the minimal error in the strain energy). (d) Distribution of the tensile stresses in the entire simulation domain and (e) in the magnified region around the crack tip. (f) A histogram using $80$ bins, showing the relative (normalized by the maximal tensile stress within the star lattice) tensile stresses and the frequencies in which they appear in the complete model.}
    \label{fig:Convergence_Fracture_Star}
\end{figure}

As the prediction of the maximal tensile stress in a beam is crucial to predicting the fracture toughness, we further assess the impact of adaptive refinement on the obtained maximum tensile stress (Fig.~\ref{fig:Convergence_Fracture_Hexagon} and Fig.~\ref{fig:Convergence_Fracture_Star}).

As before, simulations confirm the clear superiority in accuracy of the mixed-order formulation for bending-dominated topologies in coarse-grained regions (Fig.~\ref{fig:Convergence_Fracture_Hexagon}b and Fig.~\ref{fig:Convergence_Fracture_Star}b), yielding acceptable levels of error for the mixed-order QC setup across all lattice topologies. Note that, for the hexagonal topology, the value of $r_0\approx0.0017$, which yielded a minimal error in the strain energy, coincides also produces a minimal error in tensile stresses (Fig.~\ref{fig:Convergence_Fracture_Hexagon}b). However, this is not generally the case (Fig.~\ref{fig:Convergence_Fracture_Star}b) as the optimal sampling generally deviates from exact sampling (Fig.~\ref{fig:Cook_Y_Displacement}c).

Results in Fig.~\ref{fig:Convergence_Fracture_Hexagon} and Fig.~\ref{fig:Convergence_Fracture_Star} also highlight the diverse nature of stress distributions based on lattice topology -- and they underline the importance of the QC coarse-graining. While highest stresses distribute around the crack tip in the hexagonal lattice (Fig.~\ref{fig:Convergence_Fracture_Hexagon}cde), akin to the classical `lobes' observed in linear elastic fracture mechanics, the star lattice shows long shear bands, spanning almost the entire height of the modeled domain (about $900$ UCs, see Fig.~\ref{fig:Convergence_Fracture_Star}cde). Similarly, the location of highest tensile stress also varies with lattice topology (Fig.~\ref{fig:Toughness_2D}bcd).
The adaptive mixed-order QC formulation naturally refines those regions requiring higher resolution and yields acceptable errors at relatively low repUC densities (and hence at relatively low computational costs compared to the fully-resolved calculation).
Results further confirm that first-order elements are accurate when non-uniform deformation modes are sufficiently refined, in agreement with \citet{phlipot_quasicontinuum_2019} and \citet{chen_generalized_2020}. However, our mixed-order approach accurately captures non-uniform deformation, even in coarse-grained domains undergoing non-uniform deformation. 

In beam-based architected materials, the relative density $\bar{\rho}$ (i.e., the fill fraction) is commonly used in characterizing the scaling of effective material properties with density, such as the mode-I fracture toughness 
\begin{align} \label{Eq: Toughness_Scaling}
    \Bar{K}_{IC} = \frac{K_I}{\sigma_f \sqrt{l}} = D \Bar{\rho}^d
\end{align}
whose parameters $D$ and $d$ depend on the lattice architecture \citep{schmidt_ductile_2001,fleck_damage_2007,tankasala_2013_2015}. Here, $l$ is the size of the UC, and $K_I$ is the specific stress intensity factor that leads to fracture initiation. Fig.~\ref{fig:Toughness_2D} shows the above scaling law for the triangular, hexagonal, and star lattices along with the parameters $D$ and $d$ (obtained from fitting the simulated fracture toughness data points) in comparison with literature values. For the two previously studied triangular and hexagonal topologies, our results agree excellently with previous scaling laws \citep{fleck_damage_2007,tankasala_2013_2015}, while the shown data for the star lattice is new in the literature. Of course, the presented QC framework allows us to simulate larger physical samples at a fraction of the computational cost due to efficient coarse-graining of the simulation domain. For example, the first data point in Fig.~\ref{fig:Toughness_2D}a for the hexagonal lattice was obtained using only $259{,}188$ DOFs compared to $12{,}069{,}135$ DOFs in the fully-resolved case. This is a reduction in the number of unknowns of nearly $98\%$ with an accuracy greater than $99.9\%$ for the maximum tensile stress.

\subsection{Through-thickness fracture toughness of 3D beam lattices}

\begin{figure}[h!]
    \centering
\begin{tabular}{ccc}
   \includegraphics[width=0.34\textwidth]{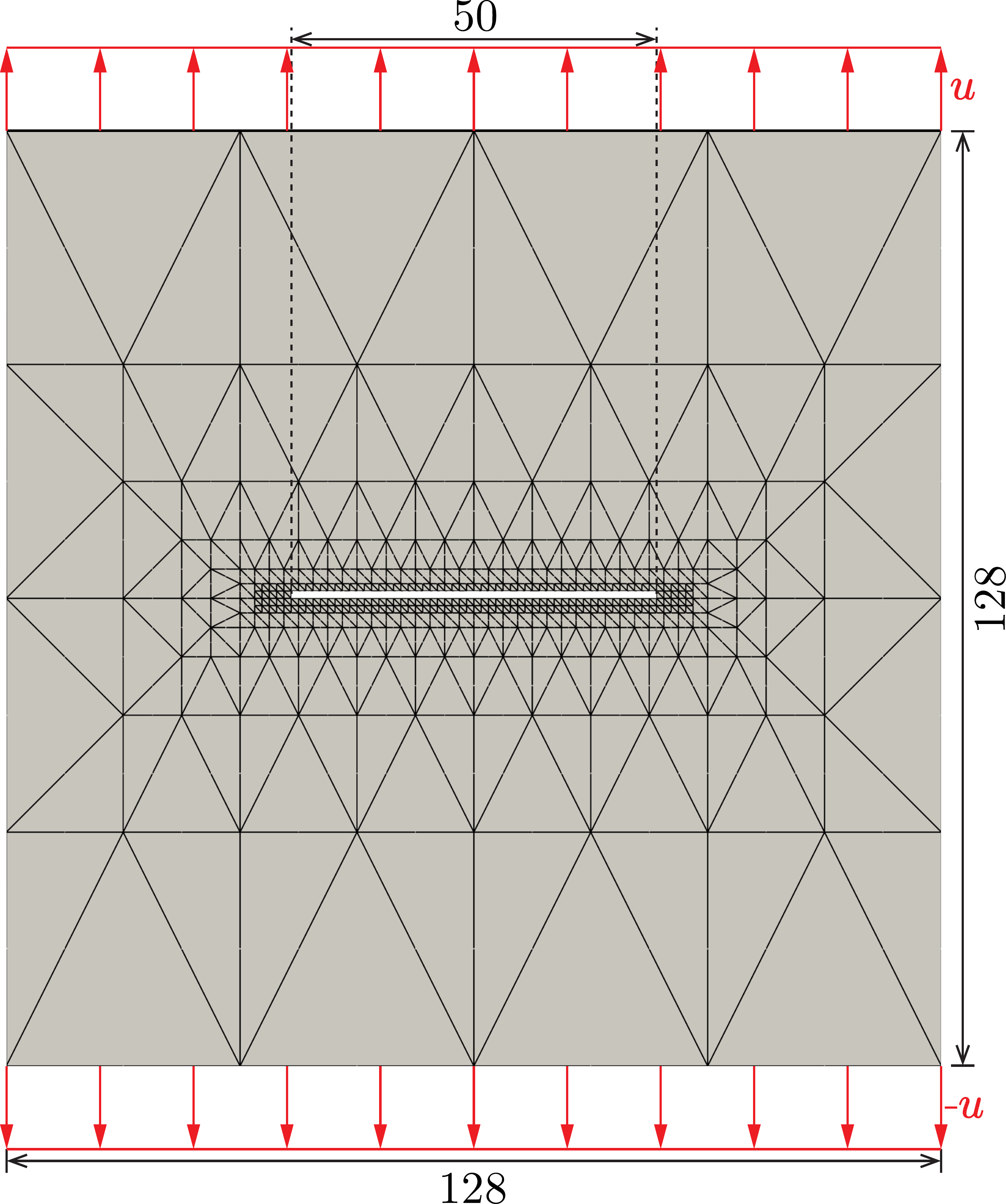} &\includegraphics[width=0.28\textwidth]{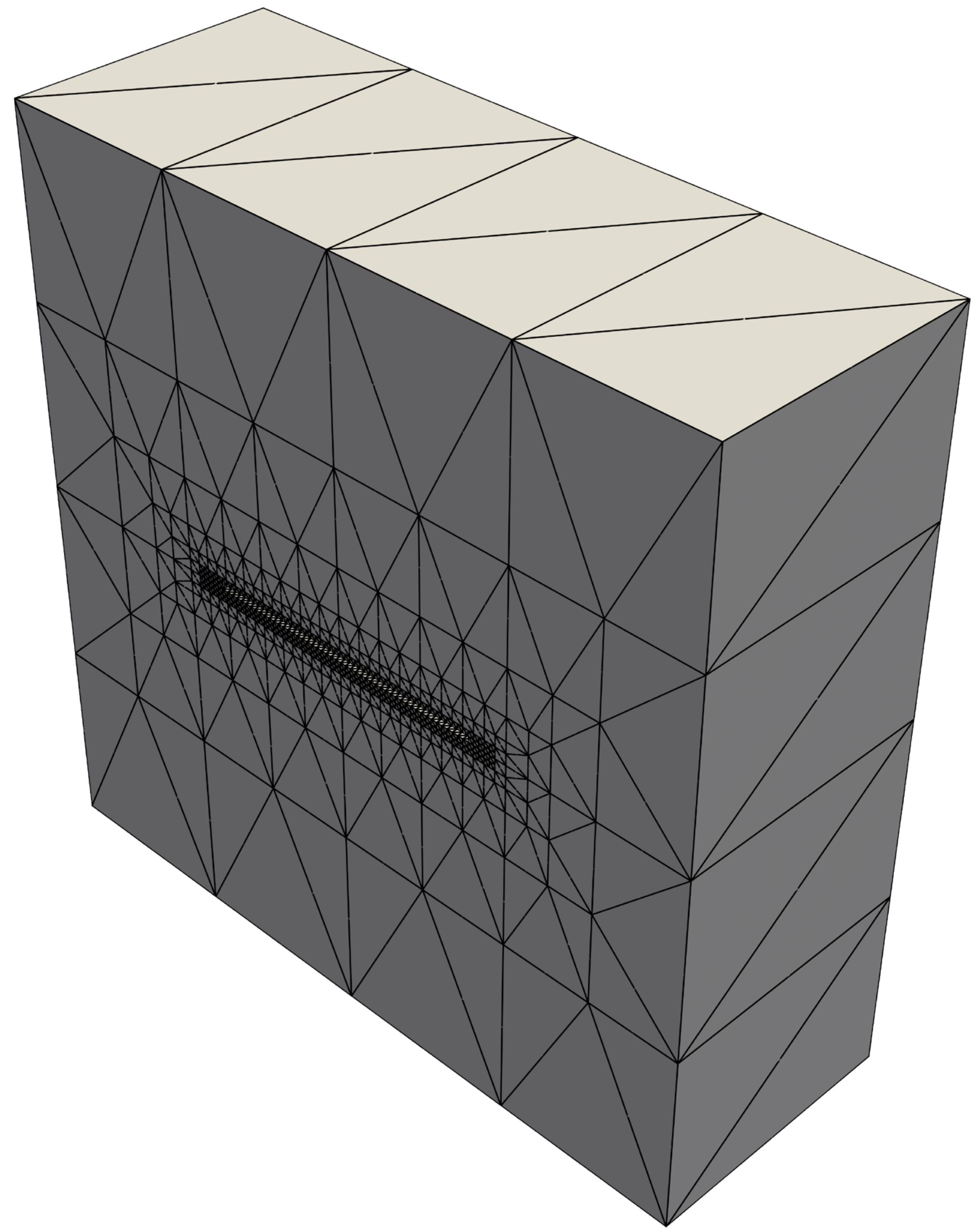}  &\includegraphics[width=0.28\textwidth]{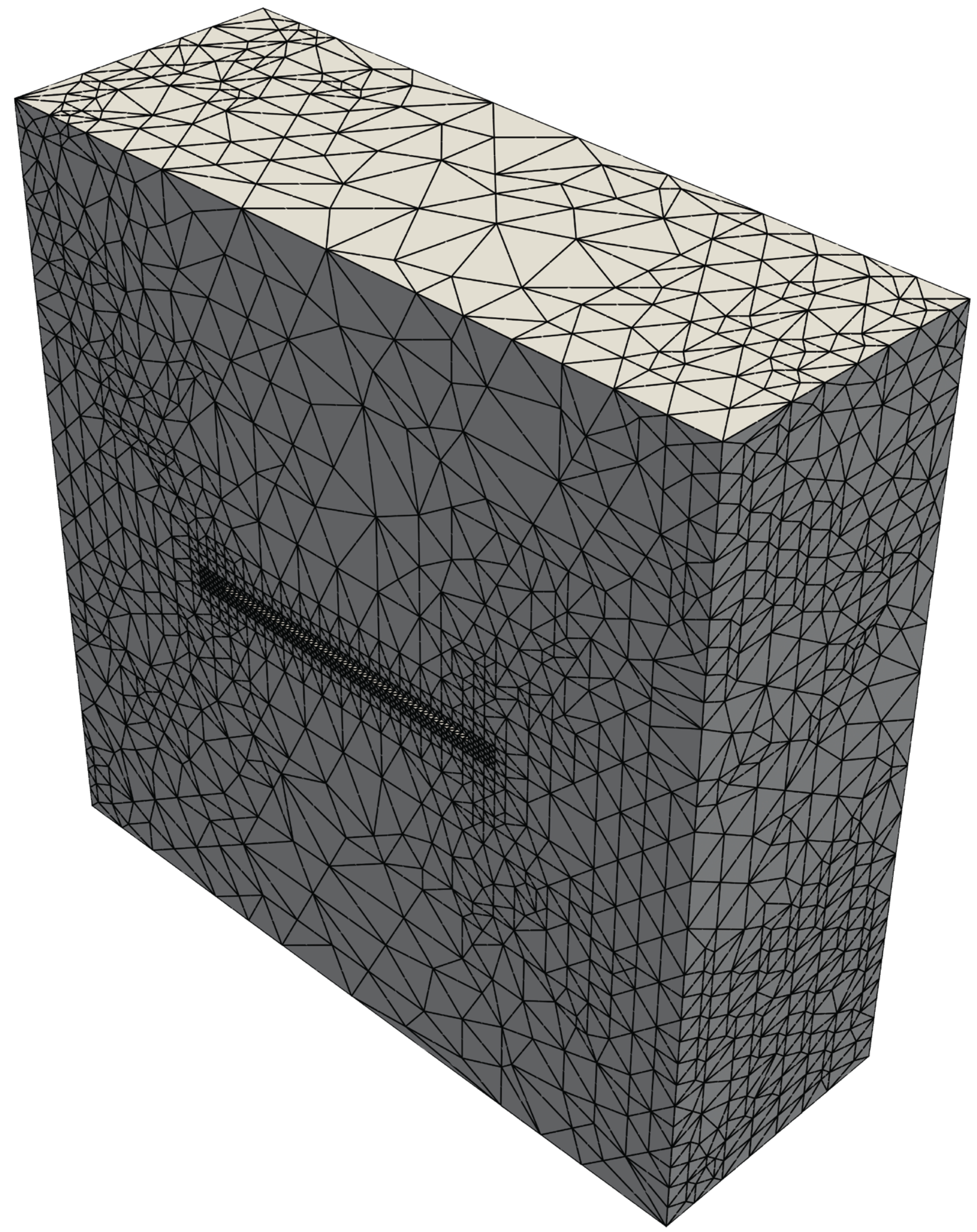} \\
(a) & (b) & (c)
\end{tabular}
    \caption{3D fracture toughness simulation setup for an \textit{octet} lattice specimen: (a) the through-thickness specimen of in-plane size $128\times128$ UCs, variable out-of-plane thickness, and a crack length of $50$ UCs, with the applied Dirichlet boundary conditions (red) at the top and bottom faces. (b) Initial QC mesh with coarsening in all three dimensions, containing $39{,}774$ repUCs. (c) Adaptively refined mesh after $8$ refinement steps, containing $157{,}321$ repUCs. The fully-resolved octet specimen contains $782{,}127$ UCs.}
    \label{fig:Refinement_3D}
\end{figure}

To fully leverage the new QC formulation, we proceed to study fracture in 3D beam lattices, where fully-resolved discrete simulations quickly become intractable.
In continuous elastic solids, the plane-strain toughness is classically considered a material property and is measured in thick specimens with a through-thickness crack \citep{astm2023standard}. In such thick samples, fracture is initiated within the solid rather than on its surface, and the obtained toughness is typically considerably lower than in thin (i.e., plane-stress) specimens. Recently, an interesting deviation from this behavior was reported for beam lattices. Specifically, \citet{shaikeea_toughness_2022} showed that the toughness of the octet lattice under plane-strain conditions exhibits little variation with specimen thickness. This was attributed to maximum beam stresses always arising on the specimen surface, so that fracture initiation always occurs on the surface for through-thickness cracks, irrespective of the specimen thickness. To investigate this behavior with our QC framework, we similarly varied the thickness of 3D octet and tetrakaidecahedral lattices, containing up to $782{,}127$ UCs in the discrete limit (with a maximum of approximately $18{,}771{,}048$ DOFs for the octet and $56{,}313{,}144$ DOFs for the tetrakaidecahedral lattice). Both simulated specimens span $128\times128$ UCs in the plane (Fig.~\ref{fig:Refinement_3D}), while the number of UCs in the out-of-plane dimension is varied. The thinnest specimen has three layers, as we need at least three UCs in the out-of-plane direction to create a second-order tetrahedral mesh. The through-thickness crack is defined by removing one layer of UCs with a width of $50$ UCs. Mimicking the conditions of \citet{shaikeea_toughness_2022}, we apply opposing uniform Dirichlet boundary conditions (a vertical displacement magnitude of $u=10^{-3}$ on the outermost nodes at the top and bottom faces) and constrain the nodes in the out-of-plane direction to achieve a discrete plane-strain condition. Therefore, in the following, the fracture toughness is related to the displacement magnitude $u$ instead of $K_I$ (as before in \eqref{Eq: Toughness_Scaling}).

\begin{figure}[b!]
    \centering
\begin{tabular}{cc}
\includegraphics[width=0.4\textwidth]{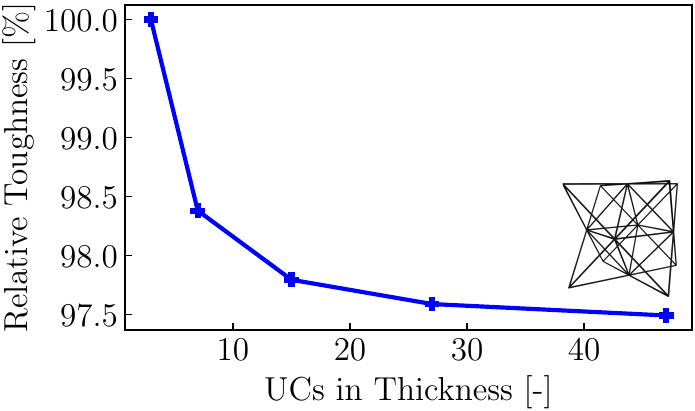}   & \includegraphics[width=0.4\textwidth]{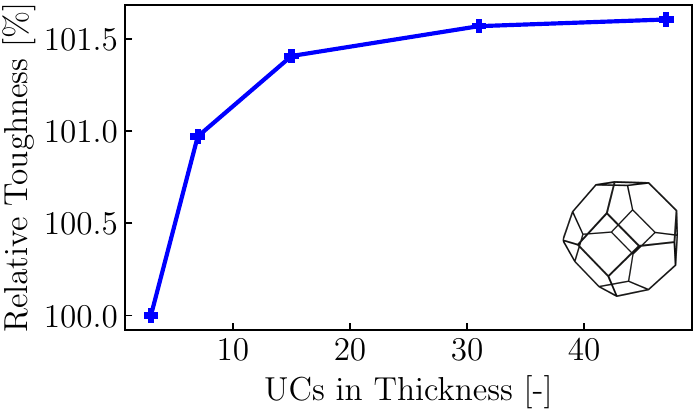} \\
     (a) & (b)
\end{tabular}
    \caption{Fracture toughness normalized by the toughness of the thinnest sample (having $3$ UCs in the out-of-plane dimension) vs.\ out-of-plane specimen thickness for (a) the octet lattice and (b) the tetrakaidecahedral lattice.}
    \label{fig:3D_01}
\end{figure}

\begin{figure}[h!]
    \centering
\begin{tabular}{c}
\hspace{-4.8cm}\includegraphics[width=0.3\textwidth]{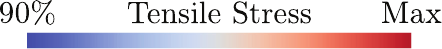}
\end{tabular}\\
\begin{tabular}{cc}
\multicolumn{2}{c}{\includegraphics[width=0.9\textwidth]{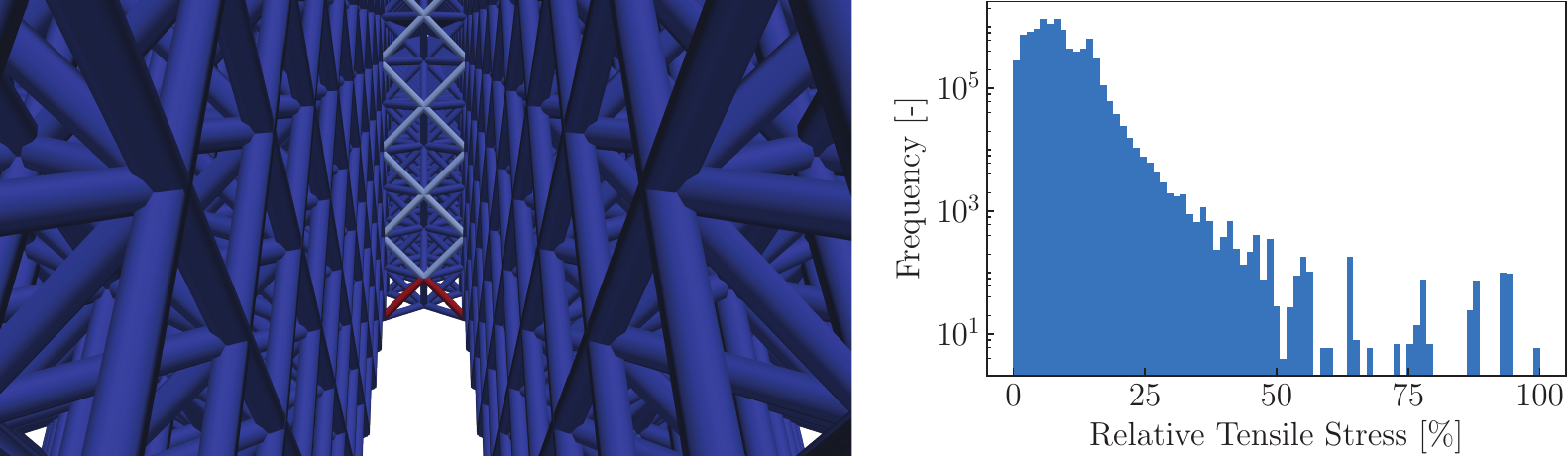}}\\
\makebox[0.52\textwidth][c]{(a)} & \makebox[0.38\textwidth][c]{(b)}\\
\multicolumn{2}{c}{\includegraphics[width=0.9\textwidth]{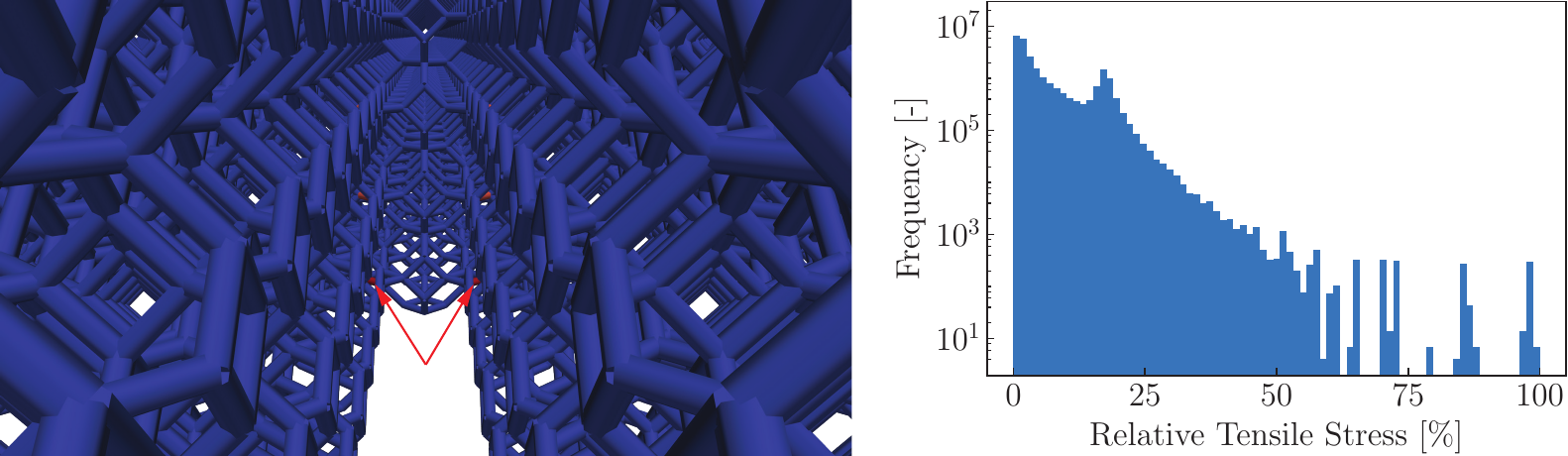}}\\
\makebox[0.52\textwidth][c]{(c)} & \makebox[0.38\textwidth][c]{(d)}
\end{tabular}    
    \caption{Stress distributions in the octet and the tetrakaidecahedral lattices with $47$ UCs in the out-of-plane dimension: the maximum tensile stresses at the bottom free surfaces are apparent for (a) the octet lattice and (b) the tetrakaidecahedral lattice. The lower end of the scale corresponds to $90\%$ of the maximal tensile stress in the respective lattices. Histograms using $80$ bins, show the relative tensile stresses (normalized by the maximum tensile stress within the respective lattice) and the frequencies at which they appear in the fully-resolved model for (b) the octet lattice and (d) the tetrakaidecahedral lattice.}
    \label{fig:3D_02}
\end{figure}

Increasing the sample thickness across the plotted range reveals an overall decrease of approximately $2.5\%$ in the octet lattice and an overall increase of approximately $1.6\%$ in the tetrakaidecahedral lattice (Fig.~\ref{fig:3D_01}ab). In both topologies, the largest change appears within the first $10$ UCs, before the values plateau at around an overall thickness of $32$ UCs. We can further confirm, that the maximum tensile stresses are always reached at the outer most layers of UCs (Fig.~\ref{fig:3D_02}ac), in agreement with prior octet observations \citep{shaikeea_toughness_2022}. The results for the tetrakaidecahedral lattice, which are new in the present study, confirm the general observation that stresses on the surface of 3D through-crack samples dominate the fracture toughness and yield a relatively small change in toughness between plane-strain and plane-stress conditions -- unlike in continuous solids, where strong differences exist. This serves as a confirmation of the theory of \citet{shaikeea_toughness_2022}. Moreover, it is interesting that, based on lattice topology, the plane-stress toughness may be higher or lower than the plane-stress value.

As the 3D nature may suggest, the highest tensile stresses are more distributed in 3D structures (Fig.~\ref{fig:3D_02}bd) than in 2D ones (Fig.~\ref{fig:Convergence_Fracture_Hexagon}f and Fig.~\ref{fig:Convergence_Fracture_Star}f), because the out-of-plane thickness gives rise to more beams around the crack tip that can carry the large tensile stresses. However, this observation breaks down at the outer surfaces, thus leading to the maximum tensile stresses visible at the free surfaces.

\section{Conclusion}
\label{sec:Conclusion}

We have presented a discrete-to-continuum coupling technique for beam lattices, which can predict, among others, the fracture toughness and crack initiation in brittle stretching- and bending-dominated periodic architected materials in 2D and 3D. Based on the quasicontinuum strategy, the method drastically reduces the number of unknowns via coarse-graining and adaptive refinement to limit full structural resolution to where it is needed (e.g., around a crack tip) while efficiently coarse-graining away from those regions. This framework enabled us to demonstrate the diverse nature of stress distributions in beam lattices as a function of lattice topology and sample dimensions, and to reliably predict their fracture toughness. Compared to all prior approaches, the mixed-order finite element interpolation (combining constant-strain and linear-strain elements) allowed us to accurately capture the highly heterogeneous or localized deformation without excessively refining the mesh up to its discrete limit. This confirms the prior conjecture that advancing from linear to quadratic elements in the coarsened domain overcomes the problems previously observed for bending-dominated beam topologies and the phenomenon introduced as stretch locking. Furthermore, the presented sampling rule with its $0^\text{th}$-order consistent weight calculation in up to three dimensions is capable to exactly capture the strain energy in lattices undergoing affine deformation only, while reducing the error below acceptable tolerances in more general deformation modes.
While the method performs excellently for all investigated topologies, one drawback is the inaccuracy of the optimal sampling rule, if compared to exact sampling. Here, a quadrature-like extension has been proposed before \citep{beex_quasicontinuum-based_2014, beex_central_2014, beex_higher-order_2015} and could be included in the future. 
A further natural extension is the modeling of inelastic material behavior and crack propagation, which will be tackled in the future based on an efficient inelastic beam formulation \citep{karapiperis_variational_nodate}. 

\section*{Acknowledgement}

The support from the European Research Council (ERC) under the European Union’s Horizon 2020 research and innovation program (grant agreement no.~770754) is gratefully acknowledged.

\clearpage
\bibliography{Library.bib}
\clearpage
\appendix
\section{Weight calculation for sampling UCs at face mid-points}
\label{app:Weights}

To calculate the weights $\omega_{s,\text{face}}$ of the face sampling UCs in 3D macroscopic elements, we leverage the Bravais coordinates of the UCs, which ensures that the three vertices forming a face $K$ of a tetrahedral element possess integer coordinates. Subsequently, we translate the macroscopic element's vertices to the origin, so they assume integer coordinates $\left\{\boldsymbol{0}, \bfY_2, \bfY_3 \right\}$, where $\bfY_2,\bfY_3 \in \mathbb{Z}^3$. Importantly, this translation preserves the count of integer points (i.e., the Bravais coordinates within the lattice $\Omega$) that intersect the face. Our aim here is to determine an explicit representation of $\Lambda=\text{span}\left(\bfY_2,\bfY_3\right) \cap \mathbb{Z}^3$. To this end, we define the normal vector
\begin{align}
    \bfn = \bfY_2 \times \bfY_3,
\end{align}
which is further divided by the greatest common divisor of its components in the $\mathcal{A}$-frame to guarantee that those become coprime.
The lattice on the triangle face is then described by the integer solutions of the homogeneous linear Diophantine equation
\begin{align}
    \bfn \cdot \bfY = 0,\, \bfY \in \mathbb{Z}^3. \label{Eq: HDLE}
\end{align}
The solutions to this equation can be written as
\begin{align}
    \bfY = c_1\bflambda_1 + c_2 \bflambda_2,\, c_1,c_2\in \mathbb{Z},
\end{align}
with $\bflambda_1,\bflambda_2\in \mathbb{Z}^3$ being non-unique basis vectors of the lattice $\Lambda$. One can show that
\begin{align}
    \bflambda_1 = \frac{\left[-n^{(2)}, n^{(1)}, 0\right]}{\text{gcd}\left(n^{(1)}, n^{(2)}\right)}
\end{align}
solves \eqref{Eq: HDLE}, and its components are coprime. To obtain the second basis vector $\bflambda_2$, we solve
\begin{align}
    n^{(1)} y_1 + n^{(2)} y_2 = \text{gcd}\left(n^{(1)}, n^{(2)}\right)
\end{align}
for $y_1$ and $y_2$ with available algorithms, such as the Extended Euclidean Algorithm \citep{mordell_diophantine_1969}. Defining
\begin{align}
    \bflambda_2 = \frac{\left[n^{(3)}y_1, n^{(3)} y_2,-\text{gcd}\left(n^{(1)}, n^{(2)} \right) \right]}{\text{gcd}\left(n^{(3)}y_1, n^{(3)} y_2, n^{(1)}, n^{(2)}\right)}
\end{align}
yields the second basis vector. Lastly, the vertices of the triangle face can be expressed by 2D integer coordinate vectors 
\begin{align}
    \bfY_i = Z_i^{(1)} \bflambda_1 + Z_i^{(2)}\bflambda_2.
\end{align}
The number of UCs within the triangle face $K$ is then calculated by applying Pick's theorem \citep{reeve_volume_1957}
\begin{align}
   \omega_{s,\text{face}} = \text{Area}\left(K \right) - \frac{b}{2} - 1,
\end{align}
where $\text{Area}\left(K\right)$ and $b$ are, respectively, the area of face $K$ when expressed in the basis $\calB = \left \{\bflambda_1,\bflambda_2 \right \}$ and the number of UCs on the boundary, (i.e., the total number of UCs on the vertices and edges). We calculate the number of UCs along the edges by utilizing \eqref{Eq: 11}, where the vertex coordinates are expressed in the $\calB$-frame.
\end{document}